\documentclass[aps,prx,11pt, tightenlines,letterpaper,onecolumn,superscriptaddress, eqsecnum,nofootinbib]{revtex4-1}

\usepackage[utf8]{inputenc}
\usepackage[english]{babel}
\usepackage[T1]{fontenc}
\usepackage{graphicx}
\usepackage{dcolumn}
\usepackage[colorlinks=true,linkcolor=blue, citecolor=cyan,urlcolor=magenta,breaklinks=true, pdftex]{hyperref}
\usepackage{dblfloatfix}
\usepackage{amsmath}
\usepackage{amsfonts}
\usepackage{amsthm}
\usepackage{enumerate}
\usepackage{bm}
\usepackage{amssymb}
\usepackage{multirow}
\usepackage{url}
\usepackage{color}
\usepackage{overpic}

\usepackage{tikz}
\usepackage{bbm}

\def\ket#1{{|{#1}\rangle}}

\usepackage{environ}
\NewEnviron{myequation}{%
\begin{equation*}
\scalebox{1.7}{$\BODY$}
\end{equation*}
}

\newcommand{\ii}{\mathrm{i}}

\begin{document}

\title{Combining Error Detection and Mitigation: A Hybrid Protocol for Near-Term Quantum Simulation}
\author{Dawei Zhong}
\affiliation{University of Southern California}
\affiliation{Lawrence Berkeley National Laboratory}
\author{William Munizzi}
\affiliation{Lawrence Berkeley National Laboratory}
\affiliation{University of California, Los Angeles}
\author{Huo Chen}
\affiliation{Lawrence Berkeley National Laboratory}
\affiliation{Harvard University}
\author{Wibe Albert de Jong}
\affiliation{Lawrence Berkeley National Laboratory}

\begin{abstract}
Practical implementation of quantum error correction is currently limited by near-term quantum hardware. In contrast, quantum error mitigation has demonstrated strong promise for improving the performance of noisy quantum circuits without the requirement of full fault tolerance. In this work, we develop a hybrid error suppression protocol that integrates Pauli twirling, probabilistic error cancellation, and the $[[n, n-2, 2]]$ quantum error detecting code. In addition, to reduce overhead from error mitigation components of our method, we modify Pauli twirling by lowering the number of Pauli operators in the twirling set, and apply probabilistic error cancellation at the end of the encoded circuit to remove undetectable errors. Finally, we demonstrate our protocol on a non-Clifford variational quantum eigensolver circuit that estimates the ground state energy of $\rm H_2$ using both \texttt{qiskit} AerSimulator and the IBM quantum processor \texttt{ibm\_brussels}. 
\end{abstract}

\maketitle

\section{\label{sec:intro}Introduction}
Quantum computing is expected to offer asymptotically faster solutions to classically difficult problems~\cite{nisq}, particularly those in quantum chemistry~\cite{qchem}, quantum simulation~\cite{qsim}, cryptography~\cite{factoring} and optimization~\cite{optim}. However, the presence and accumulation of physical errors in quantum circuits can lead to unreliable outputs, thereby preventing the realization of authentic quantum advantage. Quantum error correction codes (QECCs)~\cite{shor_code,steane_code, css_code, gottesman_thesis} are widely considered a promising approach for reversing errors and preserving high-fidelity quantum computation. Quantum error correction works by encoding information into a subspace of the full Hilbert space, known as the codespace. When an error has occurred on the stored information, the logical state is moved outside the codespace. When such a transformation has occurred, the precise deviation is identified, and a corrective operation is applied to map the logical state back into the codespace, preserving the stored logical information. Recently, several milestone results for QECCs have been experimentally verified and presented on different hardware platforms. Some notable features of these recent demonstration include codes with below fault-tolerant threshold memory~\cite{willow}, logical operations performed on multiple qubits and entangled state preparation~\cite{quera,tesseract}, teleportation of logical states~\cite{teleportation}, and experimentally realized magic state distillation~\cite{magic_quera}. 

Alongside fault-tolerant error correction, recent advancements in error mitigation have likewise been accelerating our understanding of devices in the noisy intermediate-scale quantum (NISQ) era. Notable examples of promising error mitigation techniques include zero noise extrapolation (ZNE)~\cite{zne, zne_pec}, probabilistic error cancellation (PEC)~\cite{zne_pec, pauli_lindblad}, Pauli twirling~\cite{twirling}, and measurement error mitigation (MEM)~\cite{mem_ibu}. Each of these algorithmic protocols enhances the accuracy of measured expectation values by post-processing data from an ensemble of noisy circuits. These methods are collectively referred to as quantum error mitigation (QEM)~\cite{qem}, and their utility for near-term quantum computation has been rigorously demonstrated~\cite{ibm_zne, google_purification}.

Both quantum error correction and quantum error mitigation have their respective limitations. QECCs require additional qubits and gates to protect quantum information and operations, often introducing new sources of error in the process. For this reason, successful quantum error correction demands that physical error rates in quantum devices lie below a certain threshold~\cite{threshold} to outperform the uncorrected circuit. For contemporary quantum devices, lowering physical errors to the required regime is a demanding and non-trivial task. Moreover, it is well known that Clifford unitaries are not sufficient for universal quantum computation~\cite{gottesman_thesis}. The fault-tolerant implementation of non-Clifford operations requires costly resources such as magic states. These operations often rely on resource-intensive protocols, including magic state distillation~\cite{magic_state_distill} for high-fidelity $\mathrm{T}$-states, or gate teleportation techniques for implementing gates such as the $\rm CCZ$~\cite{Nielsen_Chuang_2010}.

Meanwhile, QEM offers a practical alternative for minimizing error on NISQ devices, due to its reduced hardware overhead compared to quantum error correction. However, since QEM inherently requires averaging over many samples, the sampling costs associated with QEM grow exponentially with the number of physical error events~\cite{exp_cost}. This feature often renders contemporary QEM protocols impractical when the circuit \textit{size}, i.e. the circuit depth times the qubit number, becomes large~\cite{qem}. Considering limitations of both QECCs and QEM, it is natural to ask whether a hybrid approach could utilize the respective advantages of each while minimizing the associated overhead. Several recent attempts have been made to investigate the possibility of a hybrid combination of QEC and QEM. For example, encoded Clifford+$\mathrm{T}$ circuits can be protected by QECCs, while using PEC~\cite{qemlogicalTgate2021, qRoM} to mitigate the errors in noisy $\mathrm{T}$-state preparation. This approach reduces the hardware overhead of magic state distillation at the expense of an increased sampling cost. A similar technique, combining PEC with QEC, has been demonstrated within a fault-tolerant quantum computation (FTQC) architecture~\cite{suzukiqec+qem2022}. This integration reduces the qubit overhead required for FTQC, enabling more logical operations to be performed using the same fixed hardware resources.  

In this work, we further explore the potential of hybrid QEC/QEM approaches, and develop a framework that integrates the $[[n, n-2, 2]]$ quantum error detection code (QEDC)~\cite{iceberg, gerhard2024weakly} with Pauli twirling and PEC. QEDC is widely considered a suitable starting point for the practical implementation of stabilizer codes~\cite{422+H2, QPE+iceberg, QITE+iceberg, QAOA+ErrorDetection} since it requires only constant qubit overhead and simple post-processing. Moreover, QEDCs offer a simpler encoding scheme for single-qubit and two-qubit logical rotations~\cite{iceberg, gerhard2024weakly}, eliminating the need for compilation into Clifford+$\mathrm{T}$ circuits and avoiding circuit synthesis errors due to compilation approximation. However, there still remain a significant amount of undetectable errors in the simple, but non-fault-tolerant, encoded circuit. In an attempt to mitigation these undetectable errors, we apply Pauli twirling and PEC to eliminate remaining noise and recover noisy operator expectation values. This hybrid approach additionally reduces the sampling overhead typically required by Pauli twirling and PEC, since the use of quantum error detection lowers the level of noise to be mitigated. Accordingly, our combined approach offers mutual benefits while preserving the individual error suppression capacity for each technique. 

In this paper, we first provide a background review of quantum error detection codes (QEDC) and probabilistic error cancellation (PEC), and introduce our proposed hybrid framework in Sec.~\ref{sec:protocol}. We then review the concept of Pauli twirling and introduce our custom partial twirling protocol in Sec.~\ref{sec:QEC}. We demonstrate our combined QEDC/PEC protocol, using both simulation as well as actual IBM quantum processors, by protecting an encoded VQE circuit for $\rm H_2$ ground state energy estimation in Sec.~\ref{sec:experiment}. Finally, we summarize our results and discuss future applications in Sec.~\ref{sec:conclusion}.

\section{\label{sec:protocol}Combination of Quantum Error Detection With PEC}
In this section, we discuss how to integrate quantum error detection and quantum error mitigation. First, we will introduce the quantum error detection code in Sec.~\ref{sec:qedc} and PEC in Sec.~\ref{sec:pec}. Then we describe the error detection protocol with PEC in Sec.~\ref{sec:qedc_pec}. Finally, we explain how to determine the logical noise channel after post-selection in Sec.~\ref{sec:noise_estimation}, which is a crucial step before utilizing PEC in our protocol.  

\subsection{Quantum Error Detection Code}\label{sec:qedc}
\begin{figure}
    \centering
    \includegraphics[width=0.8\textwidth]{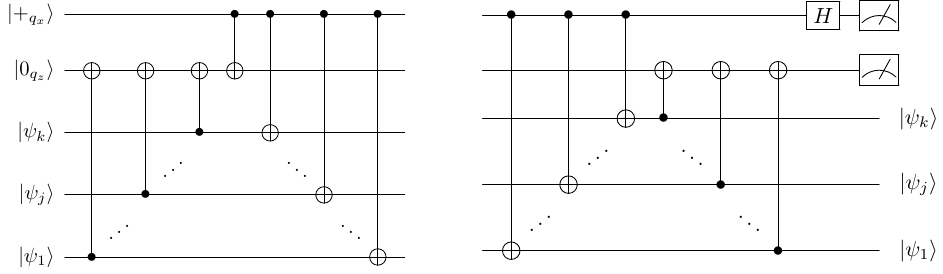}
\caption{\label{fig:encode} State preparation (left) and decoding (right) circuits for an arbitrary logical state. Note that these circuits are not fault-tolerant, but using them in this paper would not affect the main result. }
\end{figure}

The quantum stabilizer code formalism is a widely used framework for analyzing and designing quantum error-correcting codes using symmetries of quantum states. Stabilizer codes define their codespace as the joint $+1$ eigenspace of an abelian subgroup of Pauli operators, and are identified by three parameters $[[n, k, d]]$, where $n$ is the number of physical qubits, $k$ is the number of logical qubits, and $d$ is the code distance. The $[[n, n-2, 2]]$ quantum error detection code (QEDC) is a stabilizer code defined by two non-local Pauli operators, $X^{\otimes n}$ and $Z^{\otimes n}$. This family of code utilizes an even number of physical qubits $n$ to encode $k = n-2$ logical qubits with code distance $2$. If labeling $n$-physical qubits as $\{q_x, q_z, q_k, q_{k-1}, \dotsc, q_j, \dotsc, q_2, q_1\}$ where $j$ represents the $j$-th logical qubit, then the left circuit of Fig~\ref{fig:encode} is used to prepare an arbitrary logical state. Inverting the state preparation circuit gives the decoding circuit as shown in the right panel of Fig.~\ref{fig:encode}, which changes the final encoded state of $n$-qubits into the corresponding unencoded state of $(n-2)$-qubits. In addition, the decoding circuit can be used to detect errors; measuring $-1$ in any of the first two qubits suggests that the final state is corrupted and should be discarded. The logical Pauli operators on the $j$-th logical qubit is defined by $\overline{X}_{j} = X_{q_z}X_{q_j}$, $\overline{Z}_{j} = Z_{q_x}Z_{q_j}$ and $\overline{Y}_j = \ii\overline{X}_{j}\overline{Z}_j$. Here we use the notation $\overline{\,\cdot\,}$ for logical operations and $P_{q_j}$ for the Pauli operator on the physical qubit with the label $q_j$. 

To achieve universal quantum computation, a set of Clifford operations and a non-Clifford operation are required. Ref.~\cite{gerhard2024weakly} shows how to encode Hadamard, phase and $\rm CNOT$ gate into the $[[n, n-2, 2]]$ code, but implementing logical non-Clifford gates are usually difficult and resource consuming. Fortunately, the exponential operators, which is defined by $\exp(-\ii\theta P)$ with $P$ as an arbitrary $n$-qubit Pauli operator, can be encoded using quantum error detection codes by~\cite{iceberg, gerhard2024weakly}
\begin{equation}\label{eq:exp_encode}
    \overline{\exp(-\ii\theta P)} \xrightarrow{\text{ encoded into }} \exp(-\ii\theta \overline{P}),
\end{equation}
where $\overline{e^{-\ii\theta P}} = \cos \theta I - \ii\sin\theta \overline{P} = e^{-\ii\theta\overline{P}}$. Specially, the rotations on $j$-th logical qubits are given by~\cite{iceberg}
\begin{align}
    \overline{R_{X_j}}(\theta) =& \exp(-\ii\theta\overline{X}_j /2) = \exp(-\ii\theta X_{q_z} X_{q_j}/2),\\
    \overline{R_{Z_j}}(\theta) =& \exp(-\ii\theta\overline{Z}_j /2) = \exp(-\ii\theta Z_{q_x}Z_{q_j}/2).
\end{align}
In this paper, we will only use exponential maps and logical Pauli operations to implement a simple VQE algorithm (see Section~\ref{sec:experiment}).

The $[[n, n-2, 2]]$ code violates the quantum Hamming bound, so it cannot uniquely identify different errors and can only detect Pauli errors that anticommute with at least one of its stabilizer generators. The measurement of two stabilizer generators will tell whether an detectable error occurs and whether the data should be discarded. Note that Pauli errors that commute with all stabilizer generators are called logical errors. This type of noise will not trigger any of two syndrome measurements thus cannot be removed by the $[[n, n-2, 2]]$ code. 

Some coherent errors can also be detected and removed by the code. Consider an arbitrary single-qubit error $E$ which can be written as~\cite{fowler2012proof} 
\begin{equation}\label{eq:coherent_error}
    E = \begin{pmatrix}a & b \\ c & d\end{pmatrix} = \frac{a+d}{2}I + \frac{b+c}{2}X + \frac{-b+c}{2}XZ + \frac{a - d}{2}Z. 
\end{equation}
In this case, a noisy physical state $E|\psi\rangle$ can be written as a linear combination of $|\psi\rangle$, $X|\psi\rangle$, $XZ|\psi\rangle$, $Z|\psi\rangle$. If measuring stabilizer generators $X^{\otimes n}$ and $Z^{\otimes n}$, one will obtain $(+1,-1)$, $(-1, -1)$ or $(-1, +1)$ with probability $1-|(a+d)/2|^2$, which results from the term $X|\psi\rangle, XZ|\psi\rangle$ or $Z|\psi\rangle$, respectively, and can be removed by post-selection. The above procedure is still valid if $E$ is a $n$-qubit error, because the noisy state can still be expressed as a linear combination of a noisy state with detectable Pauli error $P|\psi\rangle$. 

\subsection{Probabilistic Error Cancellation}\label{sec:pec}
Probabilistic error cancelation (PEC) is designed to enhance the accuracy of the expectation value by making use of noise channel information. Assuming all gate errors are Markovian—so that errors from different layers are independent—a noisy quantum circuit with $L$-layer is
\begin{equation}\label{eq:noisy_circuit}
    \rho = \widetilde{\mathcal{U}} (\rho_{0})= \widetilde{\mathcal{U}}_{L}\circ\cdots \circ \widetilde{\mathcal{U}}_{1}(\rho_0)
    = \mathcal{N}_{L}\circ\mathcal{U}_{L}\circ \cdots \circ \mathcal{N}_{1}\circ\mathcal{U}_{1} (\rho_0),
\end{equation}
where $\widetilde{\cdot}$ denotes noisy quantity, $\rho_{0}$ and $\rho$ are the input and output state, $\mathcal{U}_l(\rho) \equiv U_l\rho U_l^{\dagger}$ and $\mathcal{N}_l$ represents the ideal operation and the associated noise of the $l$-th noisy layer $\widetilde{\mathcal{U}}_l$, respectively. The expectation value of observable $A$ measured from the above noisy experiment is
\begin{equation}\label{eq:noisy_exp}
    \langle \widetilde{A}\,\rangle = {\rm Tr}(A \,\rho) = {\rm Tr}\left[A\,\widetilde{\mathcal{U}}(\rho_0)\right]. 
\end{equation}

To estimate the noise-free expectation value $\langle \widehat{A}\,\rangle$ by PEC, one needs to identify the complete information of each $\mathcal{N}_l$, then apply the inverse of $\mathcal{N}_l$ after each layer to remove the effect of noise and finally determine $\langle \widehat{A}\,\rangle$. In practice, it is convenient to write $\mathcal{N}_l^{-1}$ as a linear combination of $\{\mathcal{O}_{\alpha_l}\}$, i.e., $\mathcal{N}_l^{-1} = \sum_{\alpha_l} \eta_{\alpha_l}\mathcal{O}_{\alpha_l}$. Here, $\{\mathcal{O}_{\alpha_l}\}$ is a set of implementable but noisy operation and $\{\eta_{\alpha_l}\}$ follows a quasi-probability distribution where $\sum_{\alpha_l}\eta_{\alpha_l} = 1$ but $\eta_{\alpha_l}$ can be negative. The quasiprobability distribution $\{\eta_{\alpha_l}\}$ can be transformed into a normal probability distribution $\{q_{\alpha_l}\}$ through $q_{\alpha_l} = |\eta_{\alpha_l}|/\gamma_l$ and $\gamma_{l} = \sum_{\alpha_l}|\eta_{\alpha_l}|\geq1$. In this case, the $l$-th layer unitary is recovered from
\begin{equation}\label{eq:pec_single}
    \mathcal{U}_{l} = \mathcal{N}^{-1}_{l} \circ\mathcal{N}_{l}\circ\mathcal{U}_l = \sum_{\alpha_l} \eta_{\alpha_l}\mathcal{O}_{\alpha_l}\circ\widetilde{\mathcal{U}}_l= \gamma_l\sum_{\alpha_l}{\rm sgn}(\eta_{\alpha_l})q_{\alpha_{l}}\mathcal{O}_{\alpha_l}\circ\widetilde{\mathcal{U}}_l
\end{equation}
where ${\rm sgn}(\eta_{\alpha_l})$ is $1$ if $\eta_{\alpha_l}>0$ and $-1$ if $\eta_{\alpha_l}<0$. The overall noiseless circuit is obtained by 
\begin{equation}\label{eq:pec_recover}
    \begin{split}
    \mathcal{U}=&\left(\mathcal{N}_L^{-1}\circ\mathcal{N}_L\right)\circ\mathcal{U}_{L}\circ\cdots \circ \left(\mathcal{N}_1^{-1}\circ\mathcal{N}_1\right)\circ\mathcal{U}_{1}\\
    =& \sum_{\alpha_{L}, \dotsc, \alpha_{1}} \eta_{\alpha_{L}}\cdots\eta_{\alpha_{1}} \mathcal{O}_{\alpha_L}\circ\widetilde{\mathcal{U}}_{L}\circ\cdots \circ \mathcal{O}_{\alpha_1}\circ\widetilde{\mathcal{U}}_{1} \\
    =& \gamma\sum_{\alpha_{L}, \dotsc, \alpha_{1}} {\rm sgn}(\eta_{\alpha_L})\cdots {\rm sgn}(\eta_{\alpha_l})q_{\alpha_{L}}\cdots q_{\alpha_{1}} \mathcal{O}_{\alpha_L}\circ\widetilde{\mathcal{U}}_{L}\circ\cdots \circ \mathcal{O}_{\alpha_1}\circ\widetilde{\mathcal{U}}_{1}.
    \end{split}
\end{equation}
where $\gamma = \prod_l \gamma_l$. Instead of calculating expectation value directly from Eq.~\eqref{eq:pec_recover}, we can creates circuits from $N$ samples of $\{\mathcal{O}_{\alpha_1}, \dotsc, \mathcal{O}_{\alpha_L}\}$ generating from probability distribution $\{q_{\alpha_1}, \dotsc, q_{\alpha_L}\}$, and estimate expectation with
\begin{equation}\label{eq:quasi_prob}
    \langle \widehat{A}\,\rangle = \frac{\gamma}{N}\sum_{k}s_k {\rm Tr}\left[A\,\mathcal{O}_{\alpha_L}\circ\widetilde{\mathcal{U}}_{L}\circ\cdots \circ \mathcal{O}_{\alpha_1}\circ\widetilde{\mathcal{U}}_{1}(\rho_0)\right]
\end{equation}
where $s_k = \prod {\rm sgn}(\eta_{\alpha_l})$ is the sign corresponding to the product of a given sample of $\{\mathcal{O}_{\alpha_l}\}$~\cite{ferracin2024efficiently}. 

Handling a general noise channel with PEC is still difficult. It is a common practice to convert a general noise channel into a stochastic Pauli noise with Pauli twirling and replace $\{\mathcal{O}_{\alpha_l}\}$ with Pauli operation $\{\mathcal{P}_{\alpha_l}\}$ as implementable noisy operations. In this paper, we will adopt this setting and utilize PEC only to cancel Pauli noise. 

\subsection{Protocol} \label{sec:qedc_pec}
\begin{figure}
    \centering
    \includegraphics[width=0.85\textwidth]{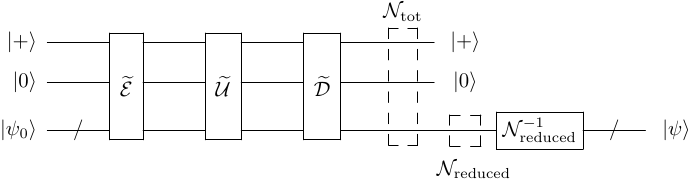}
\caption{\label{fig:protocol} The $[[n, n-2, 2]]$ error detection protocol with PEC. Here $|\psi_0\rangle$ is the input logical state. This diagram include noisy state-preparation $\widetilde{\mathcal{E}}$, noisy unitary operation $\widetilde{\mathcal{U}}$ and noisy decoding $\widetilde{\mathcal{D}}$. It also displays the ``location'' of equivalent noise channel $\mathcal{N}_{\rm tot}$ and $\mathcal{N}_{\rm reduced}$. }
\end{figure}

According to Sec.~\ref{sec:qedc}-~\ref{sec:pec}, the $[[n, n-2, 2]]$ QEDC is only able to detect all weight~$1$ and some weight~$2$ Pauli errors, while PEC can eliminate the impact of any type of noise and return a noise-free expectation. We integrate two techniques to enhance the performance of QEDC by removing logical errors with PEC. To be specific, consider a noisy encoded circuit with totally $L$-layers shown in Fig.~\ref{fig:protocol},
\begin{equation}\label{eq:noisy_encode}
    \widetilde{\mathcal{D}}\circ\widetilde{\mathcal{U}}\circ\widetilde{\mathcal{E}} (\rho_0)= \mathcal{N}_{L}\circ\mathcal{U}_{L}\circ \cdots \circ \mathcal{N}_{1}\circ\mathcal{U}_{1} (\rho_0), 
\end{equation}
where $\rho_0 = |+\rangle\langle +|\otimes |0\rangle\langle 0|\otimes |\psi_0\rangle\langle \psi_0|$ and the overall protocol includes state preparation $\mathcal{E}$, unitary operation $\mathcal{U}$, and decoding circuit $\mathcal{D}$. For $[[n, n-2, 2]]$ code, we adopt the decoding circuit in Fig.~\ref{fig:encode} for $\mathcal{D}$ and use it to convert $n$-qubit physical state into logical state. Also, the decoding circuit will be used for error detection. We will select the data if both of the first two qubits are measured as $+1$. It is possible to rewrite Eq.~\eqref{eq:noisy_encode} as
\begin{equation}
    \widetilde{\rho} = \widetilde{\mathcal{D}}\circ\widetilde{\mathcal{U}}\circ\widetilde{\mathcal{E}}(\rho_0) = \mathcal{N}_{\rm tot}\circ\mathcal{D}\circ\mathcal{U}\circ \mathcal{E}(\rho_0).
\end{equation}
where noise channel $\mathcal{N}_{\rm tot}$ is equivalent to all noise effects in the noisy encoded circuit. After error detection and post-selection, we have
\begin{equation}
    \widetilde{\rho} = \widetilde{\mathcal{D}}\circ\widetilde{\mathcal{U}}\circ\widetilde{\mathcal{E}} \xrightarrow{\text{ error detection and post-selection }} \mathcal{N}_{\rm reduced}(\rho).
\end{equation}
Here, $\mathcal{N}_{\rm reduced}(\rho)$ is the noisy quantum state after post-selection, where $\rho$ is the noiseless final state and $\mathcal{N}_{\rm reduced}$ only contains Pauli errors that cannot be removed by the QEDC. 

The implementation of our protocol is outlined below, followed by a detailed explanation of each step.

\begin{enumerate}
    \item Create a physical circuit $\mathcal{D}\circ\mathcal{U}\circ\mathcal{E}$.  
    \item Perform interleaved cycle benchmarking~\cite{PauliNoiselearnability} to learn associated noise for each of the $L$~ layers in the physical circuit. 
    \item Compute $\mathcal{N}_{\rm reduced}$ and its inverse. 
    \begin{enumerate}
        \item[3a.] Use result from Step~2 to calculate $\mathcal{N}_{\rm tot}$ by numerical simulation of error propagation. 
        \item[3b.] Delete all Pauli terms in $\mathcal{N}_{\rm tot}$ that anti-commutes with $X$ at the first qubit and anti-commutes with $Z$ at the second qubit. This step outputs $\mathcal{N}_{\rm reduced}$.
        \item[3c.] Determine the inverse of reduced noise channel $\mathcal{N}^{-1}_{\rm reduced}$.
    \end{enumerate}
    \item Execute the encoded circuit with Pauli twirling and $\mathcal{N}^{-1}_{\rm reduced}$ implemented using quasi-probability sampling. 
    \item Compute the noise-free expectation value $\langle \widehat{A}\,\rangle$ by PEC after post-selection. 
\end{enumerate}

In Step~1, we assume the physical circuit is built with single- and two-qubit Clifford gates as well as single-qubit rotation gates, which allow us to estimate the total noise channel with propagation of Pauli error. In Step~2, we assume that the noise of the same unitary operation remains unchanged across different experiments and different parts of same circuit. Therefore, the benchmarking experiment output is sufficient to describe the noise in the physical circuit and can be reused for the same unitary operation. Note that interleaved cycle benchmarking only learn Pauli noise channel, so Pauli twirling should be applied to physical circuit when using our protocol. More details of the interleaved cycle benchmarking experiment can be found in Appendix~\ref{app:cb}. 

Step~3a and 3b approximates $\mathcal{N}_{\rm reduced}$ by firstly calculate $\mathcal{N}_{\rm tot}$ then removing detectable error terms in $\mathcal{N}_{\rm tot}$. We will provide more details of these steps in Sec.~\ref{sec:noise_estimation}. In Step~4 and 5, the PEC estimator $\langle \widehat{A}\,\rangle$ is determined from
\begin{equation}\label{eq:noisefree_exp}
    \langle \widehat{A}\,\rangle = {\rm Tr}\left[ A\,\mathcal{N}^{-1}_{\rm reduced}\circ\mathcal{Q}\circ \widetilde{\mathcal{D}}\circ\widetilde{\mathcal{U}}\circ\widetilde{\mathcal{E}}(\rho_0) \right].
\end{equation}
where $\mathcal{N}_{\rm reduced}^{-1}$ is implemented by the quasi-probability sampling following Eq.~\eqref{eq:quasi_prob}, and channel $\mathcal{Q}$ denotes the post-selection prcess.

\subsection{Estimate Logical Noise for PEC}\label{sec:noise_estimation}
In this section, we are going to explain how to obtain $\mathcal{N}_{\rm tot}$ and $\mathcal{N}_{\rm reduced}$ for our protocol from circuit 
\begin{equation}\label{eq:twirl_noisy_circuit}
    \begin{split}
    \widetilde{\mathcal{D}}\circ\widetilde{\mathcal{U}}\circ\widetilde{\mathcal{E}} =& \left(\sum_{j_L} c_{j_L}\mathcal{P}_{j_L}\right) \circ\mathcal{U}_{L}\circ \cdots \circ \left(\sum_{j_1} c_{j_1}\mathcal{P}_{j_1}\right)\circ\mathcal{U}_{1}\\
    =& \sum_{j_1, \dotsc, j_L} c_{j_L}\cdots  c_{j_1}\mathcal{P}_{j_L}\circ\mathcal{U}_{L}\circ\cdots \circ\mathcal{P}_{j_1}\circ\mathcal{U}_{1},
    \end{split}
\end{equation}
where noise of each layer in Eq.~\eqref{eq:noisy_encode} becomes Pauli noise after twirling, i.e., $\mathcal{N}_l = \sum_{j_l} c_{j_l}\mathcal{P}_{j_l}$.

The first step is to get $\mathcal{N}_{\rm tot}$ via error propagation, 
\begin{equation}\label{eq:err_prop}
    \mathcal{V}\circ\mathcal{P}(\rho) = V P\rho P V^{\dagger} = (V P V^{\dagger}) V \rho V^{\dagger} (V P V^{\dagger}) = \mathcal{M}\circ\mathcal{V}(\rho),
\end{equation}
where $V$ is arbitrary unitary matrix and $\mathcal{M}(\rho) = (V P V^{\dagger})\rho (V P V^{\dagger})^{\dagger}$ is equivalent noise channel after error propagation. If $V$ is a Clifford gate, $\mathcal{M}$ will still be a Pauli channel. If $V$ is a single-qubit rotation, $V P V^{\dagger}$ will be a linear combination of Pauli operators (see Appendix~\ref{app:rot}). By repeatedly using Eq.~\eqref{eq:err_prop} to swap neighboring channels in Eq.~\eqref{eq:twirl_noisy_circuit}, we will manage to separate noise from all noise-free unitary,
\begin{equation}
    \widetilde{\mathcal{D}}\circ\widetilde{\mathcal{U}}\circ\widetilde{\mathcal{E}} = \mathcal{N}_{\rm tot}\circ \mathcal{U}_L\circ\cdots\circ\mathcal{U}_1,
\end{equation}
where 
\begin{equation}\label{eq:total_noise}
    \mathcal{N}_{\rm tot} = \sum_{j_1, \dotsc, j_L} c_{j_L}\cdots c_{j_1}\mathcal{P}_{j_L}\circ\mathcal{M}_{j_{L-1}}\circ\cdots\circ\mathcal{M}_{j_{1}}.
\end{equation}
Here $\mathcal{M}_{j_l}(\rho) = M_{j_l}\rho M_{j_l}^{\dagger}$ is equivalent noise channel of $\mathcal{P}_{j_l}$ in Eq.~\eqref{eq:twirl_noisy_circuit} after error propagation,  
\begin{equation}
    M_{j_l} = U_{L}\cdots U_{l+1}P_{j_l}U^{\dagger}_{l+1}\cdots U_L^{\dagger}.
\end{equation} 

The second step is to obtain the total logical noise channel after error detection and post-selection. It is convenient to express Eq.~\eqref{eq:total_noise} in the form of $\mathcal{N}_{\rm tot} = \sum_{k}p_k E_k\rho E^{\dagger}_k$, where the sum is over all possible $\{j_1, \dotsc, j_L\}$. $p_k = c_{j_L}\cdots c_{j_1}$ and $E_k = P_{j_L}M_{j_{L-1}}\cdots M_{j_1}$. Note that $\{c_{j_l}\}$ gives a probability distribution of noise terms in $\mathcal{N}_{l} = \sum_{j_l} c_{j_l}\mathcal{P}_{j_l}$. As a result, $\{ p_k \}$ represents a joint probability distribution. When performing error detection on $\mathcal{N}_{\rm tot}$, any error $E_k$ that is a single Pauli operator and anticommutes with a stabilizer generator will be removed. If $E_k$ is a linear combination of Pauli operators, its effects will be partially mitigated (see Appendix~\ref{app:rot} for more details). Finally, the channel for all logical noise is $\mathcal{N}_{\rm logical}(\rho) = \sum_{s}q_s E_s \rho E^{\dagger}_s$, where we renormalize $\{ p_k \}$ to get $\{q_s\}$, $\rho = |\psi\rangle\langle \psi|$ is the unencoded state after noiseless $\mathcal{D}\circ\mathcal{U}\circ\mathcal{E}$ and post-selection. In this case, the noisy expectation value is derived by
\begin{equation}\label{eq:noisy_exp_theory}
    {\rm Tr}\left[A\,\mathcal{N}_{\rm logical}(\rho) \right] = \sum_{s} q_s {\rm Tr}\left[ A\, E_s\rho E^{\dagger}_s \right].
\end{equation}

Suppose $E_s = \sum_{r_s} a_{r_s} P_{r_s}$ is a linear combination of Pauli operators, where $a_{r_s}$ is real number and $\{a_{r_s}^2\}$ gives a probability distribution, then $E_s\rho E_s^{\dagger} = \sum_{r_s, k_s} a_{r_s}a_{k_s} P_{r_s}\rho P_{k_s}$ and
\begin{equation}
    {\rm Tr}\left[ A\, E_s\rho E_s^{\dagger} \right] = \sum_{\substack{r_s, k_s\\ r_s= k_s}}a^2_{r_s}{\rm Tr}\left[ A\, P_{r_s}\rho P_{r_s} \right] +\sum_{\substack{r_s, k_s\\ r_s\neq k_s}}a_{r_s}a_{k_s}{\rm Tr}\left[ A\, P_{r_s}\rho P_{k_s} \right].
\end{equation}
The noisy expectation value under logical noise can be written as
\begin{equation}\label{eq:complete_noisy_exp}
    {\rm Tr}\left[A\,\mathcal{N}_{\rm logical}(\rho) \right] = \sum_{s, r_s} q_s a^2_{r_s} {\rm Tr}\left[ A\, P_{r_s}\rho P_{r_s} \right] + \sum_{\substack{s, r_s, k_s\\ r_s\neq k_s}} q_s a_{r_s}a_{k_s} {\rm Tr}\left[ A\, P_{r_s}\rho P_{k_s} \right].
\end{equation} 
Note that for Clifford circuit, the second term vanishes since any Pauli error will propagate through Clifford operations and become another Pauli error. For circuits with Clifford and single-qubit rotation gates, the second term is not negligible in general, it can still be addressed as a case of partial error mitigation (see Appendix~\ref{app:rot} and~\cite{zne_pec, quasiprobability} for details). In this paper, we adopt the simplest strategy by neglecting the off-diagonal terms. In this case, Eq.~\eqref{eq:complete_noisy_exp} can be rewritten as
\begin{equation}
    {\rm Tr}\left[A\,\mathcal{N}_{\rm logical}(\rho) \right]\approx {\rm Tr}\left[A\,\mathcal{N}_{\rm reduced}(\rho) \right] = \sum_{j} c_j {\rm Tr}\left[ A\, P_{j}\rho P_{j} \right]
\end{equation} 
Here, $\mathcal{N}_{\rm reduced} = \sum_{j}c_j P_j\rho P_j$ and $c_j = a^2_{r_s} \sum_{s} q_s$, and its inverse can be implemented by quasi-probability sampling. To illustrate how the idea work, we will show an example in the Appendix~\ref{app:rot} when $E_s$ is linear combination of Pauli errors. 

\begin{figure}
    \centering
    \includegraphics[width=\textwidth]{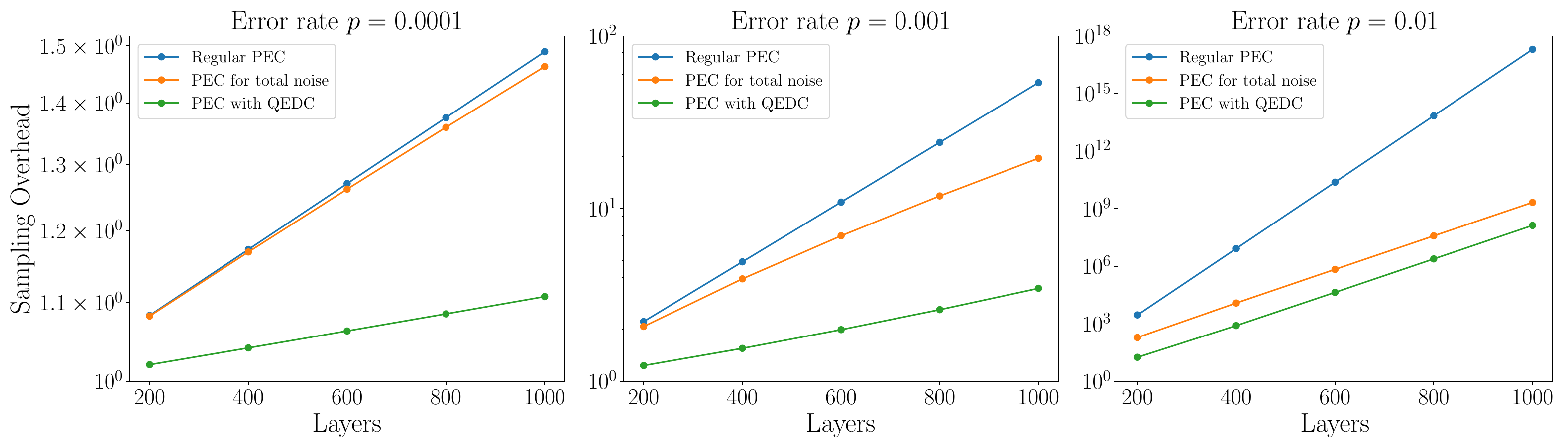}
\caption{\label{fig:overhead} Sampling overhead of three error mitigation settings: PEC for layer noise, PEC for overall noise, and combined protocol with QEDC and PEC. The first two settings are evaluated on $4$-qubits circuit, while the third one is evaluated on $6$-qubits circuit encoded with QEDC. }
\end{figure}

\subsection{Overhead}\label{sec:overhead}
The overhead of the hybrid protocol mainly comes from the sampling overhead of the PEC component, which can be evaluated by $\gamma^2$~\cite{suzukiqec+qem2022} and $\gamma$ is the parameter related to the quasi-probability distribution (see Sec.~\ref{sec:pec}). Here, we evaluate and compare $\gamma^2$ under three error mitigation settings. The first is the regular PEC protocol that cancels noise after each layer, the second is PEC that cancels estimated overall noise, and the third is the hybrid protocol with QEDC and PEC. For the first two settings, we consider $L$-layers of depolarizing noise channel with the same error rate on a $4$-qubit circuit. For the third setting, we consider $L$-layers of depolarizing noise channel but on a $6$-qubit circuit, and calculate the sampling overhead of overall noise after removing detectable Pauli errors. We do not include any types of unitary operations in our simulation for simplicity. Our results for different error rates are shown in Fig.~\ref{fig:overhead}, which indicates that PEC on estimated overall noise has a lower sampling overhead than regular PEC protocol, and PEC on encoded circuits is more efficient than PEC on unencoded circuit. More details of the calculation can be found in the Appendix~\ref{app:sampling}. 

\section{\label{sec:QEC}Partial Twirling and Partial Error-Correction}
While full fault-tolerance~\cite{Aharonov:1996yt,Calderbank:1995dw,Knill:1996ny,Fowler:2012hwn} remains a goal for the future, many NISQ error mitigation techniques~\cite{Li:2016vmf,McClean_2017,Temme:2016vkz,Cai:2022rnq} enable advanced noise suppression on contemporary hardware. In this section we introduce a protocol which utilizes a modified Pauli twirling scheme, in combination with partial quantum error correction, to minimize hardware noise on current quantum devices. Our technique incorporates an analysis of error propagation through Pauli twirling sets, selecting twirling operations in such a way as to randomize the errors toward a Pauli channel that closely matches the desired form. This protocol maximizes the utility of current error mitigation techniques, minimizing additional quantum resources while improving fidelity, and enhances the development of robust quantum devices and reliable information processing in the NISQ era.

\subsection{Pauli and Partial Pauli Twirling}

Pauli twirling~\cite{PhysRevLett.76.722,Knill:2004vlm,Kern_2005,Geller_2013} is an error mitigation technique which relies on conjugating a noisy channel by randomly-sampled Pauli gates to transform the noise into a depolarizing form. Pauli twirling is typically employed as a preliminary operation in many error mitigation schemes, enabling a simpler analysis, characterization, and eventual inversion of channel noise using auxiliary techniques such as probabilistic error cancellation (PEC). In this section we review Pauli twirling and describe the propagation of error through twirled quantum circuits~\cite{Scheiber:2024nfj}. We then introduce a modified twirling scheme, known as partial twirling, and demonstrate how it can be integrated with quantum error detection and correction. Finally, we discuss the advantages of partial twirling for improving code fidelity with reduced resource overhead, when compared to alternative approaches.

\begin{figure}[ht]
\centering
\includegraphics[width=8cm]{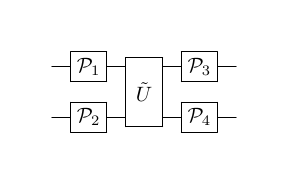}
\caption{Pauli twirling conjugates a noisy channel $\tilde{U}$ by Pauli gates $\{P_i\}$, such that the average over conjugated channels becomes a stochastic Pauli channel. Averaging over twirled circuits converts the noise $\Lambda$, in $\Tilde{\mathcal{U}}$, into a depolarizing form which can be inverted using methods such as probabilistic error cancellation.}
\label{fig:GenericTwirl}
\end{figure}

Given a unitary $\mathcal{U}$, let $\Tilde{\mathcal{U}}$ denote the imperfect realization of $\mathcal{U}$ due to noise. It is useful to express $\Tilde{\mathcal{U}}$ as the composition of the ideal $\mathcal{U}$ with a noise channel $\Lambda$, as
\begin{equation}
    \Tilde{\mathcal{U}} = \mathcal{U} \circ \Lambda.
\end{equation}
Working in the Pauli basis, we can represent $\Lambda$ as a Pauli transfer matrix which, in general, will be dense (containing many off-diagonal terms) and often non-invertible. However, through a technique known as Pauli twirling we can transform $\Lambda$ into a depolarizing channel, diagonal in the Pauli basis, by conjugating $\Lambda$ with randomly sampled Pauli gates $\mathcal{P}_i \in \Pi^n$ and averaging over all instances. The twirling set $\{\mathcal{P}_i\}$, i.e. the set of Pauli gates used for conjugation, is selected such that the action of $\mathcal{U}$ is unchanged. Fig. \ref{fig:GenericTwirl} illustrates the concept of Pauli twirling applied to a $2$-qubit gate $\Tilde{\mathcal{U}}$.

The operation shown in Fig.~\ref{fig:GenericTwirl} is repeated for all sets of Pauli gates in the twirling set $\mathcal{P} \subseteq \Pi_n$. An unbiased average is computed over all conjugated channels, building the twirled channel $\Lambda_{\mathrm{twirled}}$ as
\begin{equation}\label{TwirlingEq}
    \Lambda_{\mathrm{twirled}} = \frac{1}{|\mathcal{P}|} \sum_{\mathcal{P}_i \in \mathcal{P}} \mathcal{P}_i^\dagger \Lambda \mathcal{P}_i.
\end{equation}
The transformation of $\Lambda$ into $\Lambda_{\mathrm{twirled}}$ converts the complex noise in $\Tilde{\mathcal{U}}$, on average, into a stochastic Pauli noisy channel. As a result, the average of measured expectation values of observables is unbiased, eliminating the systematic bias introduced by accumulated coherent errors.

In this work, we combine Pauli twirling with quantum error detection (QEDC) and quantum error correction (QEC) to improve error suppression on contemporary quantum hardware~\cite{Jain_2023}. We begin by introducing \emph{partial twirling}, a modified version of the Pauli twirling protocol described above which utilizes only a subset of Pauli gates tailored to the specific QEDC/QEC scheme. Given a noisy unitary $\Tilde{\mathcal{U}}$ and a QEDC/QEC protocol capable of detecting a set of errors $\mathbb{E}$, we define the partial twirling set $\Tilde{\mathcal{P}}$ as the sets of Pauli gates which optimize the condition
\begin{equation}\label{PartialTwirlingEq}
   \min_{\Tilde{\mathcal{P}}}\left| \left| \frac{1}{|\Tilde{\mathcal{P}}|} \sum_{\mathcal{P}_i \in \Tilde{\mathcal{P}}} \mathcal{P}_i^\dagger \Tilde{\mathcal{U}} \mathcal{P}_i - \mathbb{E} \circ \mathcal{U}  \right|\right|.
\end{equation}
Eq.~\eqref{PartialTwirlingEq} formulates an optimization problem over subsets of Pauli operators, seeking those whose action on $\Tilde{\mathcal{U}}$ best aligns the resulting errors with the target error space $\mathbb{E}$ defined by the QEDC/QEC scheme. This ensures that, after twirling, propagated errors are optimally approximated by elements of $\mathbb{E}$. While the expression in Eq.\ \eqref{PartialTwirlingEq} cannot generically be minimized to zero, there are numerous cases where it does vanish, including instances where $\mathbb{E}$ consists solely of single-qubit Pauli errors, which are invariant under the twirling action. Mathematica code to generate partial twirling sets is publicly available at~\cite{githubPauli}.

When combined with partial error detection or correction, where only errors of a particular type are identifiable, partial twirling allows the choice of twirling gates to be customized to minimize undetectable errors. Moreover, the selection of twirling gates can further be adapted to the constraints of a specific hardware platform, favoring operations that are more resource-efficient or less prone to error. In this way, partial twirling provides a flexible mechanism for optimizing error suppression strategies. In the following section, we demonstrate how to integrate partial twirling with QEDC/QEC, improving the effectiveness of partial error detection and correction on NISQ hardware. We further explain how partial twirling can be implemented with reduced overhead, enabling a more efficient use of quantum resources.

\subsection{Partial Twirling with Partial Error Correction}

In the previous section, we reviewed Pauli twirling and introduced our partial twirling protocol, which strategically selects twirling gates tailored to a particular error detection/correction code or specific hardware implementation. We now demonstrate how partial twirling can be combined with error correction/detection to achieve improved error mitigation on noisy devices. We benchmark the effectiveness of partial twirling for reliably implementing IBM's native multi-qubit gates on real devices. In later sections, we extend this approach by combining partial twirling with the $[[4,2,2]]$ quantum error detection code to estimate the ground state energy of $H_2$ on noisy hardware.

\begin{figure}[ht]
\centering
\includegraphics[width=9cm]{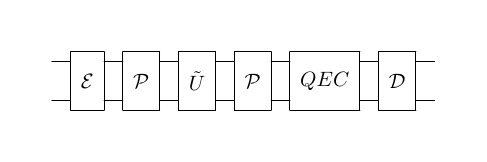}
\caption{Combination of Pauli twirling with quantum error correction (QEC). A logical state is encoded using $\mathcal{E}$, and evolved through the noisy unitary $\Tilde{\mathcal{U}}$. Twirling is implemented with operations $\mathcal{P}$ on either side of $\Tilde{\mathcal{U}}$, following which the QEC is applied. The system is decoded with $\mathcal{D}$.}
\label{fig:AbstractPartialQECFigure}
\end{figure}

A general schematic for combining twirling with error correction is given by the circuit diagram in Fig. \ref{fig:AbstractPartialQECFigure}. We begin by determining an error correction code $\mathrm{QEC}$ to implement, and subsequently encode the $k$-qubit logical state into $n$ physical qubits using an encoding unitary $\mathcal{E}$. The goal is to evolve the logical state through some noisy unitary $\Tilde{\mathcal{U}}$, representing a quantum circuit to be executed on a quantum machine, with maximum fidelity. Partial twirling is performed layer-by-layer on $\Tilde{\mathcal{U}}$, indicated by the twirling blocks $\mathcal{P}$ in Fig. \ref{fig:AbstractPartialQECFigure} (only one block is shown for clarity), with twirling gates $\mathcal{P}$ selected according to Eq.\ \eqref{PartialTwirlingEq}. After twirling and error-correction, indicated by $QEC$ in the diagram, are performed the state is passed through a decoding operation $\mathcal{D}$.
We importantly note that partial twirling does not, in general, guarantee a full conversion of the noise channel into a stochastic Pauli channel, since only a reduced subset of Pauli gates is utilized. The resulting channel depends on the specific structure of the underlying noise. Partial twirling does, however, suppress (and in in some cases completely eliminate) off-diagonal elements in the Pauli transfer matrix.

\begin{figure}[ht]
\centering
\includegraphics[width=12cm]{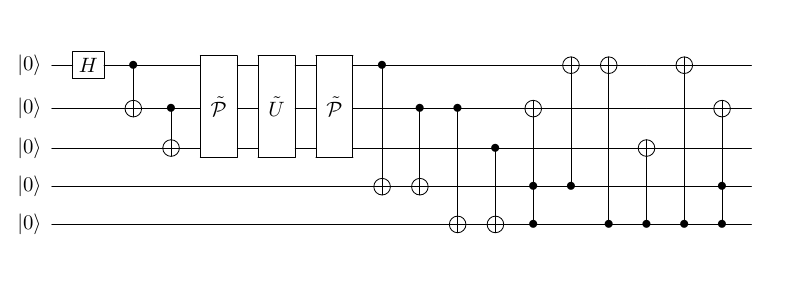}
\caption{Circuit illustrating how to combine partial twirling with the $[[3,1,1]]$ bit-flip code with two ancillary qubits. Twirling gates $\Tilde{\mathcal{P}}$ are chosen using Eq.\ \eqref{PartialTwirlingEq} to commute $X$ errors through the twirling protocol, where they can be corrected by the bit-flip code.}
\label{fig:PQEC_X_Example}
\end{figure}

When $\mathrm{QEC}$ in Fig. \ref{fig:AbstractPartialQECFigure} is a partial error correction code, only a portion of the logical code space is preserved, thereby reducing the quantum resources necessary to implement. As an example, consider the $[[3,1,1]]$ bit-flip code which can detect and reverse single-qubit Pauli $X$ error. Fig. \ref{fig:PQEC_X_Example} illustrates how to combine twirling with the bit-flip code, first encoding a $3$-qubit logical $\mathrm{GHZ}$ state into $5$ physical qubits using Hadamard and CNOT gates, and appending two ancillae to perform syndrome measurement. The state $\mathrm{GHZ}$ is to be evolved through the noisy channel $\Tilde{\mathcal{U}}$. We twirl $\Tilde{\mathcal{U}}$ using the set of gates $\Tilde{\mathcal{P}}$, determined by Eq.\ \eqref{PartialTwirlingEq}, which commute Pauli $X$ errors through the circuit where they are detected by the bit-flip code.
Fig. \ref{fig:PQEC_X_Example} provides a simple example of combining partial twirling with error correction. In more advanced implementations, partial twirling can be further employed to restrict twirling operations to gates most easily applied on a chosen hardware platform, or to avoid the dominant noise processes associated with a particular error model.

\begin{figure}[ht]
\centering
\includegraphics[width=14cm]{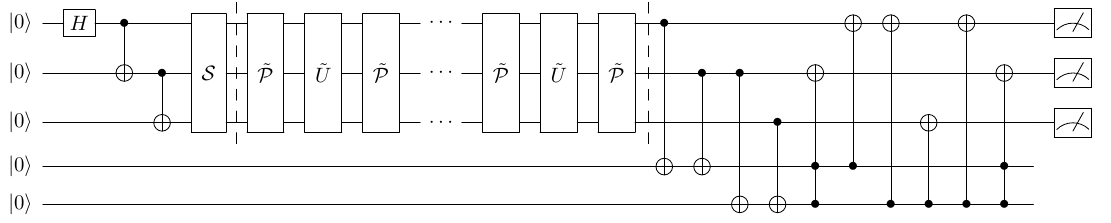}
\caption{Cycle benchmarking a noisy unitary $\Tilde{\mathcal{U}}$, using partial twirling and bit-flip code. A $5$-qubit register is initialized in a $\mathrm{GHZ}$ state on $3$ qubits, and stabilizer $\mathcal{S}$ is applied. The system is evolved through $kn$ twirled copies of $\Tilde{\mathcal{U}}$, such that $\mathcal{U}^k = \mathbb{I}$ and $n \in \mathbb{N}$. The fidelity of $\mathcal{S}$ is measured for increasing $n$.}
\label{fig:Cycle_Benchmarking_Diagram}
\end{figure}

We use cycle benchmarking to assess the performance of our combined partial twirling with $[[3,1,1]]$ code, implemented by the circuit in Fig. \ref{fig:Cycle_Benchmarking_Diagram}. As before, we begin with a $5$-qubit system and initialize the first three qubits in a $\mathrm{GHZ}$ state. We then apply an operation $S$, chosen from the stabilizer group of $\mathrm{GHZ}$ in $\Pi^3$, i.e. $S \in \{XXX,\ IZZ\}$. The sequence $\tilde{\mathcal{P}} \circ \Tilde{\mathcal{U}} \circ \tilde{\mathcal{P}}$ is applied $kn$ times, where $\mathcal{U}^k = \mathbb{I}$ and $n \in \mathbb{N}$, such that a perfect application of $\mathcal{U}$ does not modify the encoded state. After twirling $\Tilde{\mathcal{U}}$ at each layer, and evolving the encoded state through the sequence of twirled $\Tilde{\mathcal{U}}$ operations, the bit-flip code is applied and the expectation values of $S$ are measured and compared for increasing depth $n$. 

We utilize the technique outlined in Fig. \ref{fig:Cycle_Benchmarking_Diagram} to conduct cycle benchmarking experiments on IBM's multi-qubit gate set, using IBM quantum hardware. As an example, with results shown in Fig. \ref{fig:cx_cyclebench}, we benchmark the unitary
\begin{equation}
    \mathcal{U} = CX_{1,2}CX_{2,3},
\end{equation}
the composition of $2$-qubit $\mathrm{CX}$ gates acting on our encoded $3$-qubit $\mathrm{GHZ}$ state. We apply partial twirling on $\Tilde{\mathcal{U}}$, selecting twirling gates that cause Eq.\ \eqref{PartialTwirlingEq} to vanish. We compute the fidelity of stabilizers $XXX$ and $IZZ$, as a function of increasing layers of twirled $\Tilde{\mathcal{U}}$. Performances are compared between no twirling, full Pauli twirling using all $I,X,Y,Z$ combinations, and partial twirling using only $I,X,Z$ gates.
\begin{figure}
    \centering
    \includegraphics[width=0.45\linewidth]{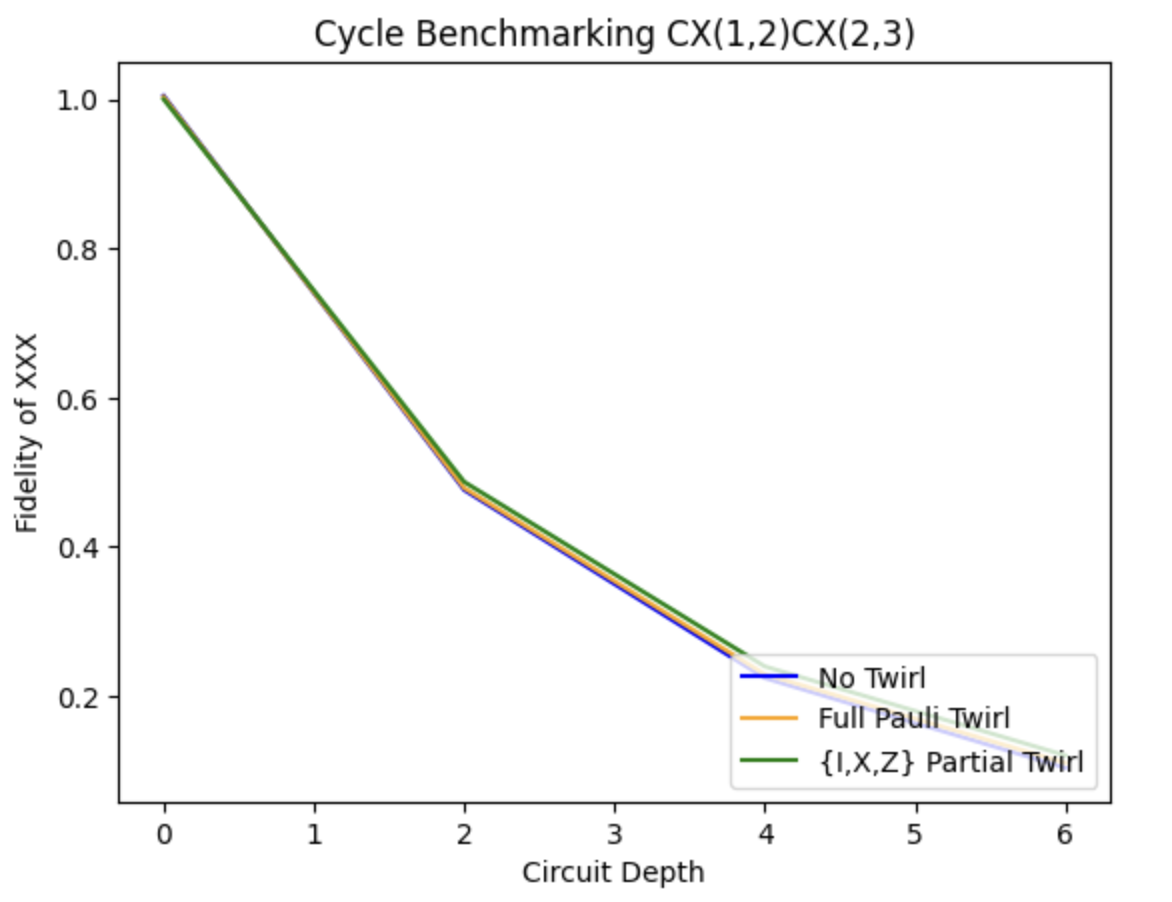}
    \includegraphics[width=0.45\linewidth]{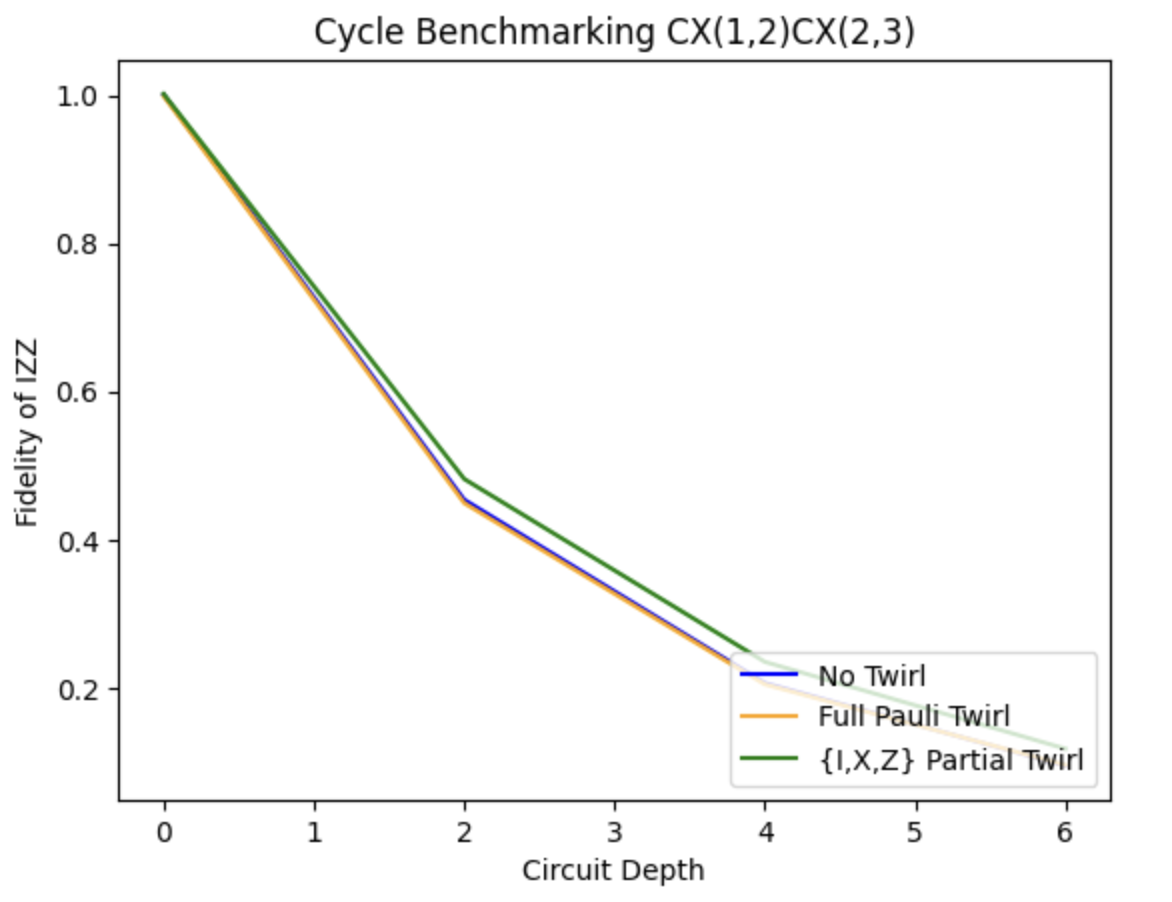}
\caption{Cycle benchmarking of $\Tilde{\mathcal{U}} = CX_{1,2}CX_{2,3}$ using twirling with bit-flip error correction. Experiment conducted on ``ibm brussels'' device. Fidelity of stabilizers $XXX$ and $IZZ$ are plotted against increasing layers of $\Tilde{\mathcal{U}}$. For partial twirling, gates are selected such that Eq.\ \eqref{PartialTwirlingEq} vanishes. Performance is compared between no twirling, full Pauli twirling with $I,X,Y,Z$ gates, and partial twirling with $I,X,Z$ gates.}
\label{fig:cx_cyclebench}
\end{figure}
We observe in Fig. \ref{fig:cx_cyclebench} that partial twirling retains higher fidelity over increasing layers of noisy $\Tilde{\mathcal{U}}$ application, compared to no twirling and full Pauli twirling. Restricting to operations that commute $X$ errors through the twirling set improves the effectiveness of the bit-flip code in within this protocol. Moreover, by restricting to a subset of Pauli gates in the twirling step, we reduce the resource overhead needed to perform the twirling protocol (see section \ref{Resources}). Cycle benchmarking results for additional IBM gates, i.e. $\mathrm{CZ}$, $\mathrm{ECR}$, and $\mathrm{iSWAP}$, are provided in Appendix \ref{app:partial}.

In this section, we demonstrated how partial twirling can be combined with partial error correction to enhance error suppression on noisy quantum hardware. We conducted an experiment, on real quantum hardware, using the $[[3,1,1]]$ bit-flip code with partial Pauli twirling. Twirling gates were selected to enforce a vanishing Eq.\ \eqref{PartialTwirlingEq}, and performance was evaluated using cycle benchmarking. In the following section, we analytically derive the logical error rate of the $[[4,2,2]]$ error detection code, under a noise channel, when twirling is applied in combination with error detection. This $[[4,2,2]]$ QEDC is then employed in Section \ref{sec:experiment} to estimate the ground-state energy of $H_2$. Finally, we highlight the resource advantages of partial twirling for near-term error mitigation on specific hardware architectures.

\subsection{Logical Fidelity of Twirling with Error Detection}

When combining randomized compiling with quantum error detection/correction, an important consideration is the effect of channel noise on the logical error rate.
In this section we analytically derive the logical fidelity of the $[[4,2,2]]$ error detection code~\cite{Vaidman_1996,Rains:1997je} under a noise process. We consider both the application of Pauli twirling and partial twirling, and likewise derive their respective effects on preserve the logical fidelity of our code.

We first consider a unitary noise process, characterized by an over-rotation about the $Z$ direction, instantiate by the channel
\begin{equation}\label{ZChannel}
    \mathcal{E}\left(\rho\right) \equiv R_Z\left(\omega\right)\rho R_Z\left(-\omega\right), \qquad \textnormal{with} \quad R_Z\left(\omega \right) = \cos(\omega/2)I + i\sin(\omega/2)Z
\end{equation}
On a system of $n$ qubits, $\mathcal{E}\left(\rho\right)$ acts independently on each qubit, and therefore the effect of this noise channel on an encoded logical state $\overline{\rho}$ is 
\begin{equation}\label{TotalZChannel}
    \mathcal{E}^{\otimes n}\left(\overline{\rho}\right) \equiv R_Z^{\otimes n}\left(\omega\right)\overline{\rho} R_Z^{\otimes n}\left(-\omega\right).
\end{equation}

The $[[4,2,2]]$ error detection code is a $4$-qubit CSS code, with $2$-qubit logical codeword basis given by
\begin{equation}\label{Codewords}
\begin{split}
     \ket{\overline{00}} &= \frac{1}{\sqrt{2}}\left(\ket{0000} + \ket{1111} \right),\\
     \ket{\overline{01}} &= \frac{1}{\sqrt{2}}\left(\ket{0011} + \ket{1100} \right),\\
     \ket{\overline{10}} &= \frac{1}{\sqrt{2}}\left(\ket{0101} + \ket{1010} \right),\\
     \ket{\overline{11}} &= \frac{1}{\sqrt{2}}\left(\ket{0110} + \ket{1001} \right),\\
\end{split}
\end{equation}
The stabilizer group $\mathcal{S}$ that leaves $\{\ket{\overline{00}}, \ket{\overline{01}}, \ket{\overline{10}}, \ket{\overline{11}}\}$ invariant is generated the two Pauli strings

\begin{equation}\label{StabGen}
    \mathcal{S} = \langle XXXX,\ ZZZZ \rangle.
\end{equation}

We determine the effect of randomized compiling on code performance by computing the contribution to the code's logical fidelity~\cite{Iyer_2022,Jain_2023} from the noise process $\mathcal{E}^{\otimes n}\left(\overline{\rho}\right)$. For a fixed choice of operator basis $\{E_m\}$, e.g. the Pauli basis, we can express a quantum channel $\mathcal{E}(\rho)$ as the sum
\begin{equation}\label{ChiEquation}
    \mathcal{E}(\rho) = \sum_{m,n} \chi_{m,n}E_m \rho E_n^\dagger,
\end{equation}
where $\chi$ denotes the Chi-matrix representation~\cite{Wood:2011zvw}, a reshaped presentation of the Choi matrix, for the channel. Using Eq.\ \eqref{ChiEquation}, we construct the Chi-matrix operator for the logical noise channel $\mathcal{E}^{\otimes n}\left(\overline{\rho}\right)$, which we denote $\overline{\chi}$. Given an error correction code capable of detecting a set of errors $\mathcal{E}_C$, the diagonal elements of $\overline{\chi}$ contain the correction probabilities associated with each $E \in \mathcal{E}_C$. The total probability of correctable errors $p_c$ is then computed~\cite{Iyer_2022,Jain_2023} as the indexed sum over diagonal elements in $\overline{\chi}$, specifically
\begin{equation}\label{ProbCorrectable}
    p_c = \sum_{E \in \mathcal{E}_C} \chi_{E,E}.
\end{equation}
Consequently, the total probability of uncorrectable errors $p_u$ is given by $p_u = 1-p_c$. The value of $p_u$ relates to the randomized benchmarking infidelity $r$ as
\begin{equation}\label{ProbUncorrectable}
    p_u = r - \sum_{\substack{E\in\mathcal{E}_C\\
    E \neq \mathbb{I}}} \chi_{E,E}.
\end{equation}
For a noise channel decomposable into Pauli noise processes, $p_u$ computes the logical infidelity $\overline{r}$ of the code under the noisy process~\cite{Iyer_2022,Jain_2023}. Thereby we calculate the average logical infidelity $\overline{r}$ of our $[[4,2,2]]$ code, under $\mathcal{E}^{\otimes n}\left(\overline{\rho}\right)$, using 
\begin{equation}\label{LogicalInf}
    \overline{r} = p_u - \sum_{\substack{E,E'\in \mathcal{E}_C,\ E\neq E'\\
    s(E) = s(E'),\ \overline{E} = \overline{E'}}} \chi_{E,E'},
\end{equation}
where $E$ and $E'$ are different errors in $\mathcal{E}_C$ that produce equivalent syndromes, i.e. $s(E) = s(E')$, and admit the same overall effect on the logical data, i.e. $\overline{E} = \overline{E'}$. 

Since Pauli and partial twirling transform the matrix $\chi$, the mitigative effects of twirling on the logical infidelity of our code can be derived from Eq.\ \eqref{LogicalInf}. As a function of rotation parameter $\omega$, as in Eq.\ \eqref{TotalZChannel}, the logical fidelity of the $[[4,2,2]]$ code under $\mathcal{E}^{\otimes n}\left(\overline{\rho}\right)$ decays as 
\begin{equation}\label{UntwirledInf}
   \overline{r}_{\mathrm{Untwirled}} =  1 - 6\cos \left(\frac{\omega}{2} \right)^6 \sin \left(\frac{\omega}{2}\right)^2.
\end{equation}
Similarly computing $\overline{r}$ for the case where full Pauli twirling is applied, we have a logical infidelity
\begin{equation}\label{FullTwirledInf}
   \overline{r}_{\mathrm{Full \, Twirling}} =  1 - 4\cos \left(\frac{\omega}{2} \right)^6 \sin \left(\frac{\omega}{2}\right)^2.
\end{equation}
Finally, for the case where we apply partial twirling using sets composed of $\{I,X,Z\}$ we find that the logical infidelity of $[[4,2,2]]$ becomes
\begin{equation}\label{ParTwirledInf}
   \overline{r}_{\mathrm{Par. \, Twirling}} =  1 - 4\cos \left(\frac{\omega}{2} \right)^6 \sin \left(\frac{\omega}{2}\right)^2 - \frac{1}{288}\csc \left(\frac{\omega}{2} \right)^4 \sin \left( \omega\right)^6.
\end{equation}

\begin{figure}[h]
    \begin{center}
        \begin{overpic}[width=11cm]{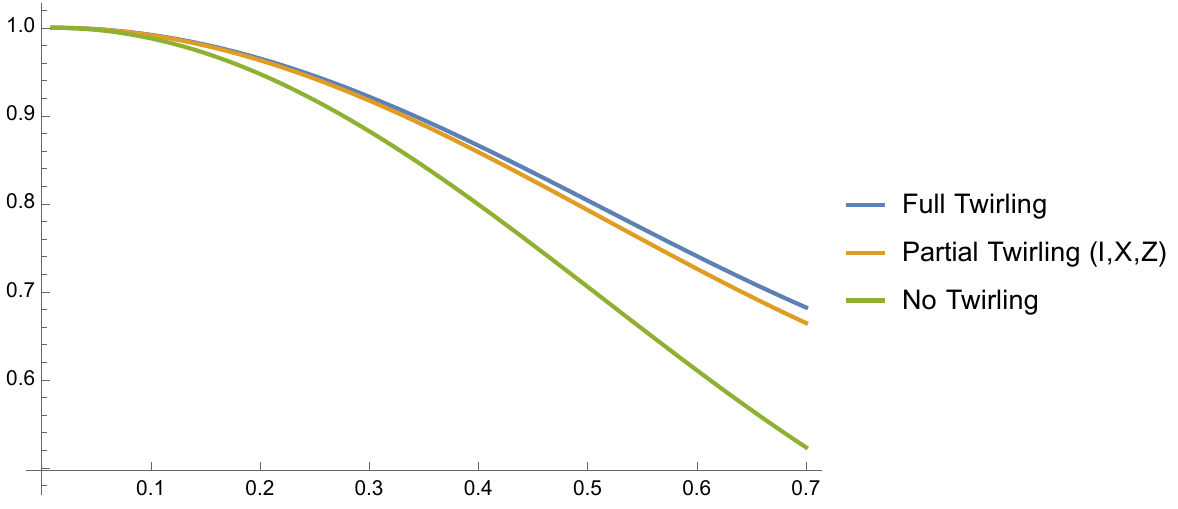}
            \put (-3,25) {$\bar{r}$}
            \put (40,-2) {$\omega$}
        \end{overpic}
    \caption{Logical infidelity of the $[[4,2,2]]$ code under the noisy process given by Eq.\ \eqref{TotalZChannel}, plotted against increasing error $\omega$. We consider the case when no twirling is applied on the noisy channel, when full Pauli twirling is applied, and when partial twirling using the reduced gate set $\{I,X,Z\}$ is applied.}
    \label{LogInf}
    \end{center}
\end{figure}

Figure \ref{LogInf} illustrates the three expressions for logical infidelity given by Eqs.\ \eqref{UntwirledInf}--\eqref{ParTwirledInf}, plotted against increasing rotation error $\omega$.
As shown in Figure \ref{LogInf}, both partial and full twirling significantly improve the $[[4,2,2]]$ code fidelity in the presence of the noise channel $\mathcal{E}^{\otimes n}\left(\overline{\rho}\right)$. Full Pauli twirling, where twirling sets composed of all Pauli gates are sampled and averaged over, performs best since the $\chi$ matrix is rendered diagonal after the twirling protocol is applied. While partial twirling does not fully diagonalize the $\chi$ matrix of the error channel, it nevertheless significantly reduces the magnitude of off-diagonal elements. This reduction offers mitigating effects, with a significant reduction in sampling during the twirling stage.

In this section we explained how to compute the logical infidelity of a quantum error correcting code, under a noisy process. We evaluated the logical infidelity of the $[[4,2,2]]$ error detection code subject to $Z$ rotation noise, comparing performance with and without Pauli twirling applied to the noise channel. We demonstrated that partial twirling is sufficient to retain code fidelity in the presence of noise, though it performs slightly worse than full Pauli twirling. In the following section, we highlight resource advantages of partial twirling and identify additional benefits of our hybrid error mitigation strategy.

\subsection{Advantages and Overhead of Partial Twirling}\label{Resources}

A concern when utilizing any error mitigation scheme is the associated resource overhead, which impacts both the efficiency and feasibility of practical implementation. While contemporary error mitigation methods are designed to improve circuit fidelity, this improvement often comes with exponential sampling cost~\cite{exp_cost}. Accordingly, minimizing resource overhead is essential for any viable error mitigation strategy. For the case of Pauli twirling, one source of overhead arises from the insertion of single-qubit Pauli gates before and after the noisy unitary to be twirled. The cost of applying these single-qubit operations varies by platform, and depends heavily on hardware design and physical qubit architecture. That being said, single-qubit gates are extremely fast and high-fidelity (with error rates often around $10^{-6}$) in many current hardware architectures, making the cost of implementing Pauli gates effectively negligible.

One significant advantage of partial twirling is realized on hardware platforms with a strong error bias, where certain error types occur more frequently than others. A specific example is the cat qubit~\cite{GuillaudCohenMirrahimi2023,AutonomousQEC_SqueezedCat2023,FaultTolerantSpinCat2024, Xu2022_KerrCat}, implemented using superconducting microwave cavities, where devices are engineered to intentionally suppress errors along a particular direction. In this setting, partial twirling allows users to restrict to gates which can be realized with minimal error. When combined with error correction, this approach further enables efficient error suppression with reduced resource overhead.

Another advantage of partial twirling arises in platforms where certain gates are easier to implement than others. For example, in many superconducting transmon qubit systems, a Pauli $X$ gate (which can be executed via a simple microwave pulse) is typically very fast and reliable, whereas a Pauli $Y$ gate might require a different calibration and could be slightly less accurate; Pauli $Z$ rotations are often implemented as frame shifts (software updates) with effectively no physical error. In such a scenario, partial twirling makes it possible to favor the high-fidelity gates in the twirling set. One could twirl using only ${I, X, Z}$ operations, for instance, avoiding $Y$ gates if they are known to be error-prone. A smaller twirling set composed of higher-fidelity operations reduces the chance that the twirling procedure itself introduces additional errors~\cite{Cai_2019}. 

When incorporating Pauli twirling within an error mitigation protocol, such as probabilistic error cancellation (PEC), one primary concern is the cost of sampling overhead. Sampling overhead describes the cost associated with executing numerous circuits, or \textit{shots}, to accurately characterize the desired channel. Recall that Pauli twirling relies on sampling numerous twirled circuits to convert a noisy operation into the form
\begin{equation}\label{TwirlingEq}
    \Lambda_{\mathrm{twirled}} = \sum_{i}^N c_i \mathcal{P}_i^\dagger \Lambda \mathcal{P}_i,
\end{equation}
where $N$ counts the number of twirled circuits being sampled over and $c_i$ the appropriate coefficient for each twirled instance. When all Pauli gates are used for twirling, and all twirled circuits are sampled, $N$ scales as $4^n$ for an $n$-qubit experiment. 

Conversely, since partial twirling only ever selects a subset of $\Pi^n$ to twirl with, the number of shots $N$ is reduced and the sampling cost to implement error mitigation is lessened. Recent publications have shown that reduced twirling sets, such as those implemented in our partial twirling protocol, are capable of providing a comparable improvement to average logical fidelity when compared to twirling with the complete group of operators~\cite{Cai_2019}. We importantly note, however, that many recent error mitigation schemes will choose to only sample a fixed $N$ number of circuits, regardless of the gate set used in the Pauli twirling step. The value of $N$ is typically chosen such that an average over $N$ twirled circuits approximates a stochastic Pauli channel, though the channel is not exact. For the experiments performed in Section \ref{sec:experiment} of this paper, we too impose a cutoff on the number of circuits sampled, instead of sampling over the full group generated by our chosen twirling operators.

In this section, we introduced an error suppression protocol that integrates Pauli twirling with quantum error correction and detection. We introduced the technique of partial twirling, where a subset of twirling operations is selected to best align with the capabilities of a chosen error correction scheme, as well as the particular constraints of a given hardware platform. As a concrete example, we implemented partial twirling in combination with the $[[3,1,1]]$ bit-flip code on five qubits, selecting twirling operations which commute with propagating Pauli $X$ errors. Using this hybrid method, we performed cycle benchmarking on IBM’s native two-qubit gates to evaluate performance in the presence of real device noise. We then analytically derived the logical infidelity of the $[[4,2,2]]$ error detection code, under $Z$ rotation noise, comparing the mitigative effects of full Pauli twirling and partial twirling. Our analysis highlighted the resource advantages gained by restricting the twirling set to a subset of Pauli gates, as well as by restricting error correction to a subspace of the logical code space. In the next section, we apply our hybrid protocol with the $[[4,2,2]]$ code to estimate the ground-state energy of $H_2$ on noisy quantum hardware, demonstrating improved accuracy and reduced sampling cost when compared to other contemporary methods.

\section{\label{sec:experiment}Experiment} 
In this section, we will demonstrate the hybrid protocol of QEDC and PEC with partial Pauli twirling on a $4$-qubit VQE circuit that estimates the ground-state energy of $\rm H_2$. We first provide a detailed setting of the algorithm in Sec.~\ref{sec:problem}, then verify the protocol on the simulator in Sec.~\ref{sec:simulator}, and finally show both expectation values and the ground state energy evaluated from the IBM quantum processor \texttt{ibm\_brussels} in Sec.~\ref{sec:ibm}.

\begin{figure}
    \centering
    \includegraphics[width=\textwidth]{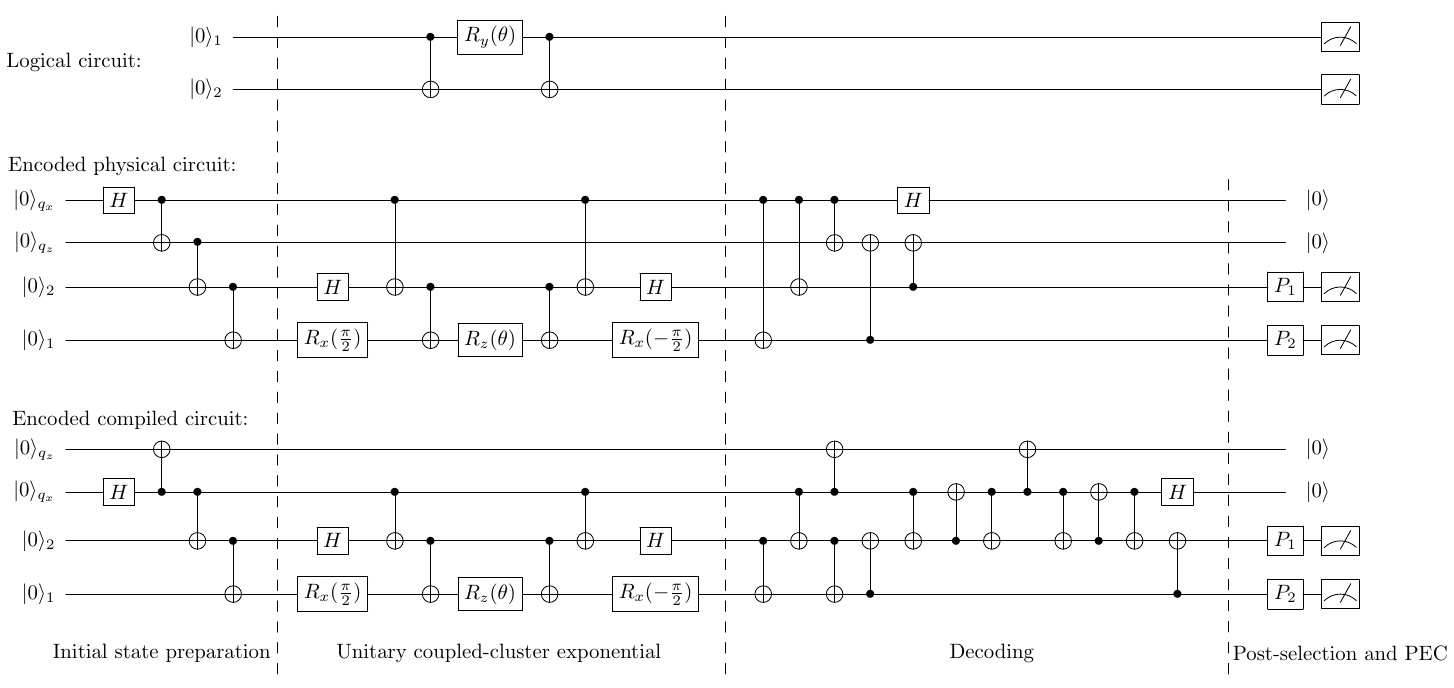}
\caption{\label{fig:experiment_circ} Quantum circuits that implement the VQE algorithm for $\rm H_2$ ground state energy estimation. Encoding the logical circuit (top) using the $[[4,2,2]]$ code leads to the physical circuit (middle) with initial state preparation $\mathcal{E}$, unitary operation $\mathcal{U}$ and decoding $\mathcal{D}$. The compiled circuit (bottom) is transformed from the physical circuit by swapping the first two-qubit and replacing long-range $\rm CNOT$ gates with equivalent neighboring $\rm CNOT$ gates.}
\end{figure}

\subsection{Experimental Setting}\label{sec:problem}
Our experiment adopts the smallest QEDC, which maps an arbitrary two-qubit state into a subspace of four-qubit Hilbert space as follows.
\begin{equation}\label{eq:encoding}
    \begin{split}
    |\overline{00}\rangle =& \frac{1}{\sqrt{2}}(|0000\rangle + |1111\rangle),
    |\overline{01}\rangle = \frac{1}{\sqrt{2}}(|0110\rangle + |1001\rangle), \\
    |\overline{10}\rangle =& \frac{1}{\sqrt{2}}(|0101\rangle + |1010\rangle),
    |\overline{11}\rangle = \frac{1}{\sqrt{2}}(|0011\rangle + |1100\rangle).
    \end{split} 
\end{equation}
The stabilizer generators of the $[[4,2,2]]$ code are $X_1X_2X_3X_4$ and $Z_1Z_2Z_3Z_4$. The logical Pauli operators for encoded state in eq.~\eqref{eq:encoding} are given by
\begin{equation}\label{eq:logical_pauli}
\overline{X}_1 = X_{2}X_{4}, \overline{X}_2 = X_{2}X_{3}, \overline{Z}_1 = Z_{1}Z_{4}, \overline{Z}_2 = Z_{1}Z_{3}, \overline{Y}_1 = i\overline{X}_1\overline{Z}_1, \overline{Y}_2 = i\overline{X}_2\overline{Z}_2.
\end{equation}

We follow the settings in Ref.~\cite{422+H2} for our experiments. The transformed Hamiltonian of $\rm H_2$ molecule is given by
\begin{equation}
    H = g_1 + g_2 Z_1 + g_3Z_2 + g_4Z_1Z_2 + g_5X_1 X_2.
\end{equation}
The VQE algorithm solves for the ground state by searching for the parameter $\theta$ in the unitary coupled-cluster (UCC) ansatze, $|\psi(\theta)\rangle = e^{-i\theta Y_1X_2/2}|00\rangle$, which minimizes the energy
\begin{equation}\label{eq:energy}
    E(\theta) = \langle \psi(\theta)|H|\psi(\theta)\rangle = g_1 + g_2\langle Z_1\rangle_{\theta} + g_3\langle Z_2\rangle_{\theta} + g_4\langle Z_1Z_2\rangle_{\theta} + g_5\langle X_1X_2\rangle_{\theta}.
\end{equation}
The quantum circuit that implements the UCC ansatze is shown at the top of Fig.~\ref{fig:experiment_circ}.

Three components are required to construct the encoded circuit: initial state preparation $\mathcal{E}$, encoded unitary $\mathcal{U}$ and decoding $\mathcal{D}$. In our experiment, the first part will be a circuit that constructs a $4$-qubit GHZ state as encoded $|00\rangle$ state, and the last part is a $4$-qubit version of decoding circuit in Fig.~\ref{fig:encode}. The second part circuit should implement non-Clifford logical operation $e^{-\ii\theta Y_1X_2/2}$, which is encoded into $e^{-\ii\theta Z_1X_2Y_4/2}$ according to eq.~\eqref{eq:exp_encode} and eq.~\eqref{eq:logical_pauli}. All above circuit blocks are combined to form a complete four-qubit physical circuit, as shown in the middle of Fig.~\ref{fig:experiment_circ}. The bottom circuit of the same figure is a compiled version of physical circuit after mapping onto a linear configuration of qubits. Although this compilation may not be fully optimized, it would not affect the demonstration and performance of our protocol.

\begin{figure}
    \centering
    \includegraphics[width=\textwidth]{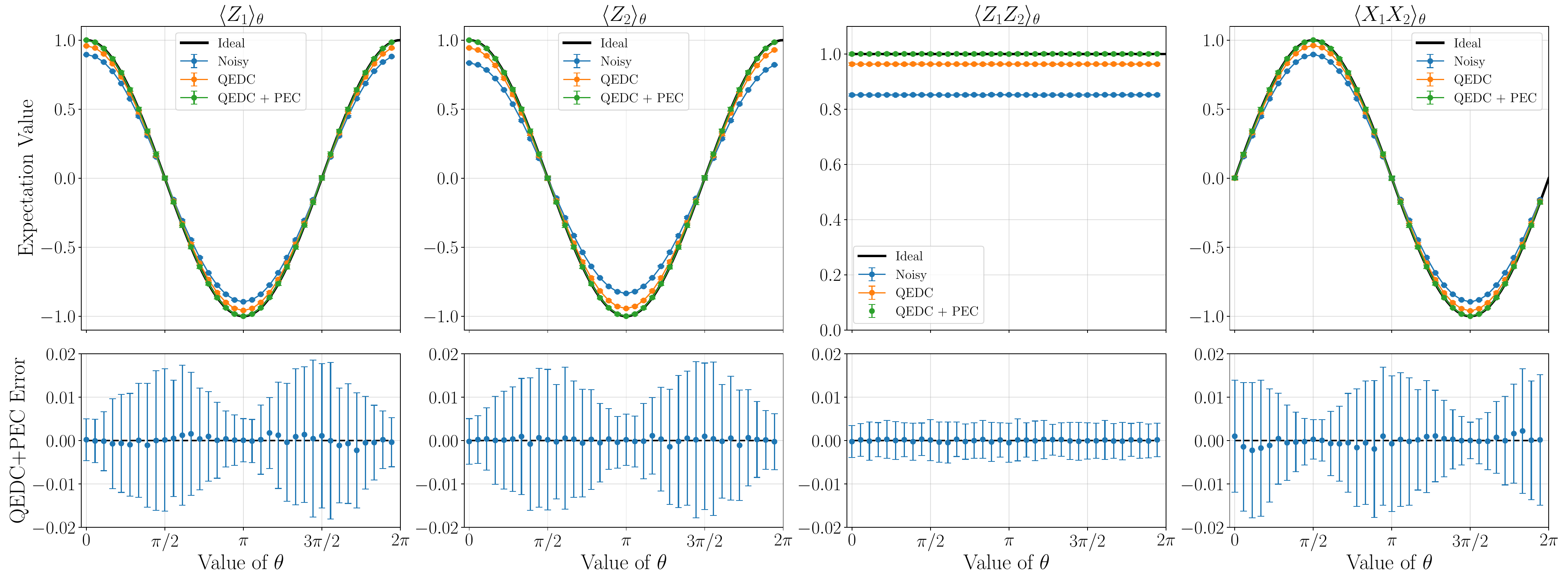}
\caption{\label{fig:simulator_exp} Simulation results of compiled circuit without any error suppression, with QEDC and with our combined protocol. The top panels show the averaged expectation values plus error bars at two standard deviation. The bottom panels give their differences from the exact values. }
\end{figure}

\subsection{Experiment on Simulator}\label{sec:simulator}
To demonstrate the protocol in Sec.~\ref{sec:qedc_pec}, we simulate the compiled circuit in Fig.~\ref{fig:experiment_circ} using \texttt{qiskit} AerSimulator. We employ a $2$-qubit depolarizing noise for all $\rm CNOT$ gates with error rate $p=0.01$, which is consistent with the ECR error rate of selected qubits for the real device experiment (see Sec.~\ref{sec:ibm}). We do not perform Pauli twirling in our simulation since the noise is already a Pauli channel. 

There are two essential post-processing steps to perform the hybrid protocol: estimation of $\mathcal{N}_{\rm reduced}$ and recovery of the noise-free expectation value from the data after post-selection. For the first step, we follow Sec.~\ref{sec:noise_estimation} and firstly estimate the overall noise channel $\mathcal{N}_{\rm tot}(\rho,\theta) = \sum_{k}p_k E_k(\theta)\rho E_k^{\dagger}(\theta)$. Here, $E_k$ can be a single Pauli operator or a linear combination of Pauli operators, and $\theta$ is the parameter of the UCC ansatz. Then we set $p_k=0$ if every term in $E_k(\theta)$ contains $X$ or $Y$ on any of the first two qubits, and renormalize the coefficients for $\mathcal{N}_{\rm logical}$, which represents the remaining noise after applying QEDC. Finally, we discard the off-diagonal terms in $\mathcal{N}_{\rm logical}$ and renormalize the coefficients again to obtain the reduced noise channel $\mathcal{N}_{\rm reduced}$. In this case, $\mathcal{N}_{\rm reduced}$ is a two-qubit Pauli channel and its inverse, $\mathcal{N}^{-1}_{\rm reduced}(\theta) = \sum_{j=1}^{16} c^{\rm inv}_j(\theta) \mathcal{P}_j$, can be computed from the inverse of Pauli transfer matrix~\cite{introGST} and the Walsh-Hadamard transformation. 

\begin{figure}
    \centering
    \includegraphics[width=0.45\textwidth]{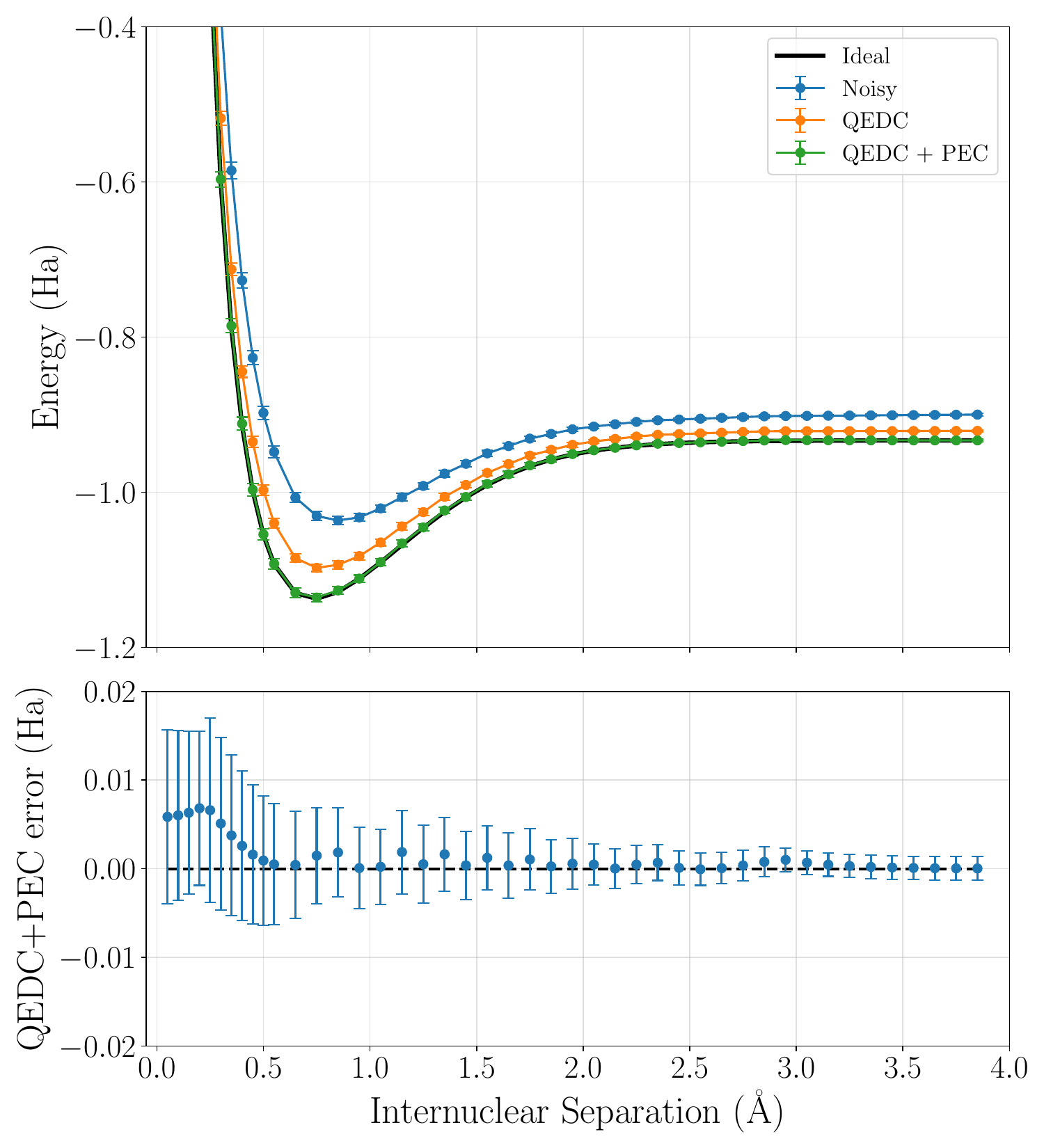}
    \caption{\label{fig:simulator_groundstate} Simulation result of $\rm H_2$ ground state energy as a function of internuclear distance. The top panel gives the result from different error suppression techniques, while the bottom panel only show the difference between exact value and the value from the protocol integrating QEDC and PEC. }
\end{figure}

For the recovery step, we estimate the noise-free expectation value via
\begin{equation}\label{eq:experiment_A}
    \left\langle \widehat{A}(\theta)\,\right\rangle = \sum_{j=1}^{16} c_j^{\rm inv}(\theta) {\rm Tr}\left\{A \,\mathcal{P}_j\circ\mathcal{N}_{\rm reduced}[\rho(\theta)] \right\},
\end{equation}
where $\rho(\theta)$ is the two-qubit final state after decoding and post-selection. Note that in our paper, $\langle \widehat{A}(\theta)\,\rangle$ is not obtained from quasi-probability sampling; instead, we construct $16$ compiled circuits with different $P_1\otimes P_2$ and sum over expectation values measured from each circuit after performing post-selection. We found that the final results show no significant difference between quasi-probability sampling and the direct summation.

Our simulation results are shown in Fig.~\ref{fig:simulator_groundstate}, where each data point represents the average of expectation value from $100$ independent simulations. Error bars indicate two standard deviations across experiments. Fig.~\ref{fig:simulator_groundstate} shows the potential energy surface of $\rm H_2$ under different separation of two hydrogen atoms. Here, we adopt coefficients of Eq.~\eqref{eq:energy} from Ref.~\cite{coefficient_H2} (See Appendix~\ref{app:coeffient}). In order to compare the performance of different error suppression techniques, we also give the expectation values and the ground state energy obtained from data without any post-processing (labeled as ``noisy'') as well as data after selecting measurements with $+1$ for the first two qubits (labeled as ``QEDC''). Our results indicate that the hybrid protocol can effectively recover the noisy expectation values to the ideal one when all physical noise is Pauli noise.  

As mentioned in Sec.~\ref{sec:noise_estimation} and Appendix~\ref{app:rot}, the off-diagonal term in the $\chi$-matrix of logical noise channel $\mathcal{N}_{\rm logical}$ could affect the expectation value. We examine the logical noise channel in the simulation and conclude that under depolarizing error rate $p=0.01$ for all $\rm CNOT$ gates, the second term in Eq.~\eqref{eq:complete_noisy_exp} will bring at most $\pm 0.001$ bias to $\langle Z_1\rangle, \langle Z_2\rangle, \langle X_1X_2\rangle$ and $\pm 10^{-5}$ bias to $\langle Z_1Z_2\rangle$. This is negligible compared to the error bars of expectation values in Fig.~\ref{fig:simulator_exp}. 
\begin{figure}
    \centering
    \includegraphics[width=\textwidth]{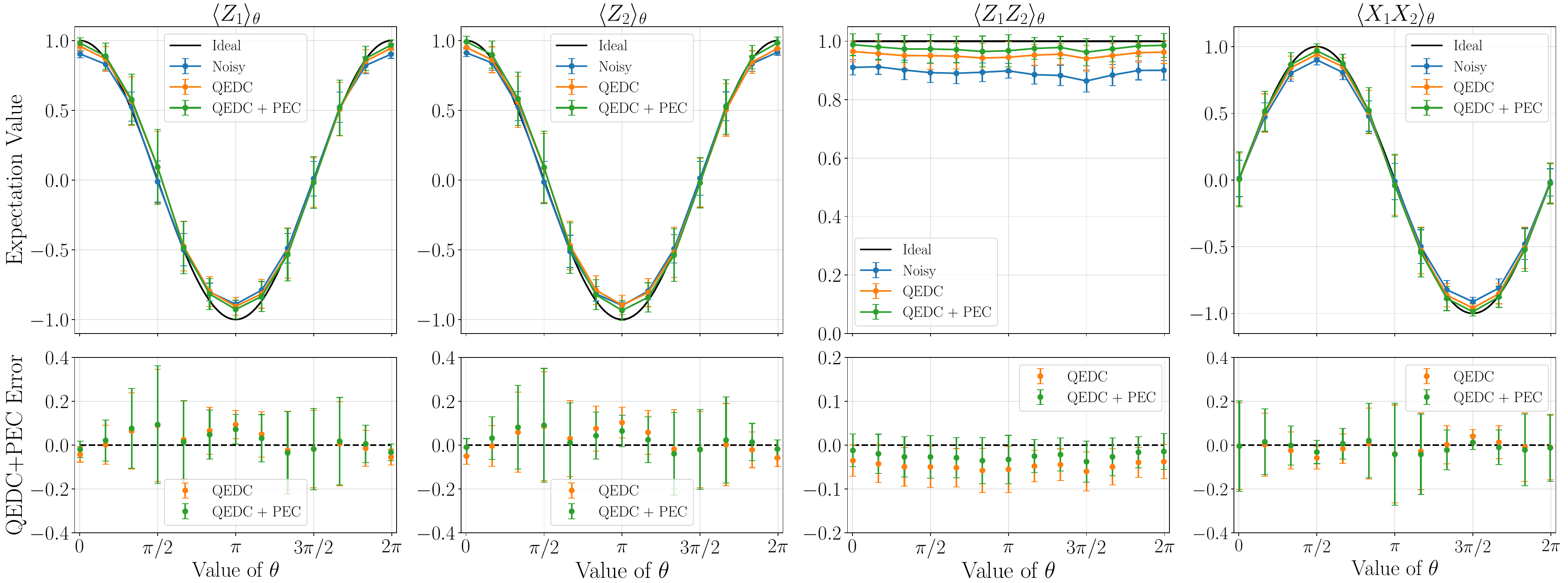}
    \caption{\label{fig:partial_exp} The expectation value obtained from \texttt{ibm\_brussels}. All measurements are calculated after readout error mitigation. The result is consistent with the simulation result in Fig.~\ref{fig:simulator_exp}. }
\end{figure}

\begin{figure}
    \centering
    \includegraphics[width=0.45\textwidth]{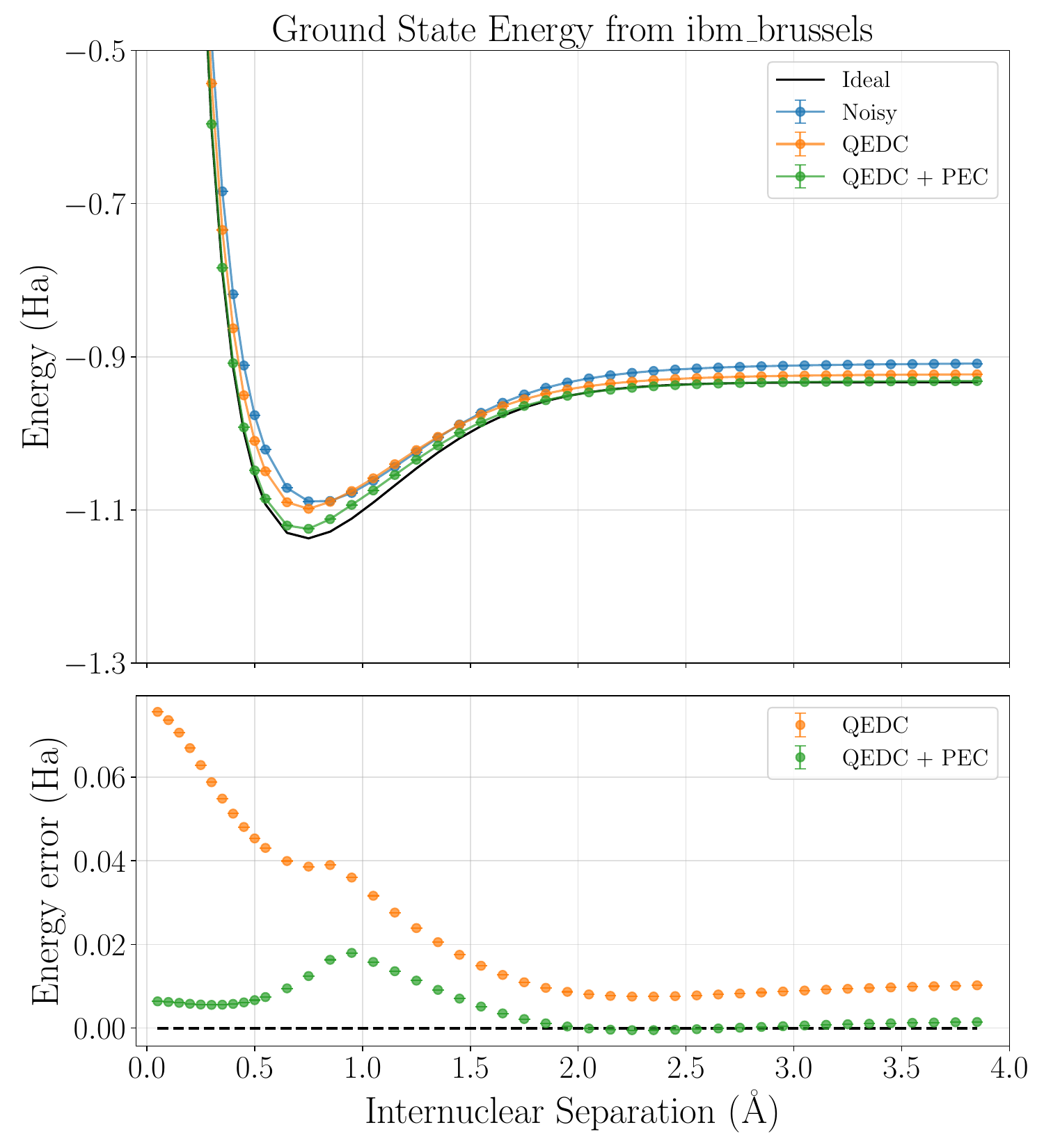}
\caption{\label{fig:partial_gstate} $\rm H_2$ ground state energy estimated from real device experiments.  }
\end{figure}

\subsection{Experiment on IBM Quantum Processors}\label{sec:ibm}
We further demonstrate the integrated protocol using the VQE experiment together with partial Pauli twirling on the IBM Quantum \texttt{ibm\_brussels} superconducting processor. First, we test compiled circuit without any error suppression on different configuration of qubits, and determine which of them is not significantly affected by hardware noise thus return reasonable result. As a consequence, we select a linear chain of qubits labeled as \texttt{19-18-14-0} to measure $\langle Z_1\rangle_\theta$, $\langle Z_2\rangle_\theta$, $\langle Z_1Z_2\rangle_\theta$, and $\langle X_1X_2\rangle_{\theta}$. We then perform cycle benchmarking~\cite{CB2019, PauliNoiselearnability} to learn the Pauli fidelities $\boldsymbol{f}$ of $\rm ECR$ gates (native gate of \texttt{ibm\_brussels}) acting on neighboring qubit pairs in the above two sets of qubit layout. The Pauli fidelities $\boldsymbol{f}$ should be converted into Pauli error rates $\boldsymbol{c}$ using the Walsh-Hadamard transformation for further usage (see Appendix~\ref{app:cb} for details and results). Finally, we apply partial Pauli twirling on all $\rm ECR$ gates using twirling set $\{IY, IZ, YY, YZ, ZI, ZY, ZZ\}$, and run compiled circuits with light optimization (i.e., transpiler optimization level as $1$) and other settings the same as those for simulation. 

Fig.~\ref{fig:partial_exp} plots the expectation values of the Hamiltonian under purely partial twirling, QEDC after partial twirling and the protocol integrating partial twirling, QEDC and PEC. We found that the readout error is a major source of noise that cannot be reduced by above error mitigation techniques. Hence, we employ the iterative Bayesian unfolding~\cite{mem_ibu} with iteration as $2$ to mitigate readout errors for all observables with minimum uncertainty. The error bars are estimated from the combination of variance of each term in Eq.~\eqref{eq:experiment_A}, and are given at two standard deviations around the data points. Our results are consistent with the simulation results in Fig.~\ref{fig:simulator_exp}, and the hybrid protocol can further suppress the noise and nearly recover noise-free expectation values as expected. The small discrepancy between experiment and ideal noise-free values may come from several sources, such as imperfect Pauli twirling, inaccurate noise estimation, and other hardware noise (single-qubit gate noise, crosstalk, etc.) that are not included in our noise model.  

Finally, we interpolate the $13$ data points in Fig.~\ref{fig:partial_exp} with a cubic spline to obtain smooth expectation-value curves as a function of $\theta$, and evaluate the potential energy surface of $\rm H_2$ by minimizing Eq.~\eqref{eq:energy} over $\theta$ at each fixed internuclear distance. Our results under different error reduction protocol are shown in Fig.~\ref{fig:partial_gstate}. Here, the internuclear distance at which the energy is minimized under QEDC or the hybrid protocol matches that of the ideal curve (around $0.75$\AA), whereas the distance of minimum from the noisy data is slightly shifted to the right (between $0.75{-}0.85$\AA). Also, the energy minimum from the hybrid protocol is $-1.124(7)$~Ha, which is very close to the exact value $-1.137$~Ha, while the minimum from noisy data and QEDC data are $-1.088(8)$~Ha and $-1.098(5)$~Ha, respectively. All these results suggest that the accuracy of VQE algorithm can be improved by our hybrid protocol. The main sources of bias in the energy estimate are the choice of interpolation method and parameters, as well as the limited number of data points.

To compare the sampling overhead of different approach, we compute $\gamma^2$ for overall error suppresion process using noise learning data from cycle benchmarking, and obtain $\gamma^2 \approx 1.19, 1.82, 1.86$ for our protocol, PEC at the end of circuit, and PEC after each layer, respectively. The results indicate that our protocol has lower sampling overhead than purely PEC, which is consistent with discussion in Sec.~\ref{sec:overhead}.

\section{\label{sec:conclusion}Discussion and Conclusion}
While fault-tolerant quantum computing remains an long-term goal for the eventual success of quantum computation, a hybrid error suppression approach offers a promising path to near-term application. In this paper, we develop a hybrid protocol which integrates Pauli twirling and probabilistic error cancellation with the $[[n, n-2, 2]]$ quantum error detection code. The partial Pauli twirling scheme we proposed in Sec.~\ref{sec:QEC} requires less than all $4^n$ Pauli operators. In contrast to Ref.~\cite{Cai_2019} which also construct a smaller twirling set, our method is designed for the usage together with quantum codes since it manipulates noise channel such that it becomes more detectable or correctable. Moreover, the probabilistic error cancellation in Sec.~\ref{sec:protocol} is modified to inverse noise after error detection instead of noise of each layer (see e.g., Ref.~\cite{pauli_lindblad}), where the remaining noise is evaluated via Pauli error propagation. Recently, there have been some efforts in the literature that estimate and remove the overall noise of Clifford circuit/subcircuit~\cite{error_estimate_pec, pec_sv} with probabilistic error cancellation. However, these papers do not discuss how Pauli error propagates through single-qubit rotation gates and how to recover expectation under the impact of resulting non-stochastic noise, which is a crucial step in our protocol as described in Sec.~\ref{sec:qedc_pec}-\ref{sec:overhead}. 

This framework offers a mutual advantage to quantum error detection code and quantum error mitigation. From the perspective of quantum error detection, both Pauli twirling and probabilistic error cancellation boost its performance because they eliminate undetectable errors in the encoded circuit. Meanwhile, numerical analysis in our paper shows that the sampling cost of probabilistic error cancellation is decreased when it is only used to remove undetectable errors in the encoded circuit. To demonstrate its utility on contemporary noisy hardware, we apply our protocol on a VQE circuit for $\rm H_2$ ground state energy estimation and observe a significant improvement in the measured expectation value. This is the first quantum hardware implementation of a hybrid error suppression protocol based on PEC. 

It is worth noting that even for a $n$-qubit Clifford circuits, probabilistic error cancellation for estimated overall noise is classically expensive, since it requires $O(4^n)$ classical memory to store complete information of $n$-qubit Pauli noise channel. To address this challenge, one should consider a sparse noise model for large quantum circuits that contains only the leading order errors. For example, a first-order noise model for individual $n$-qubit layer with $k$ two-qubit gates includes $15k$ types of error if only one of the $k$ gates is faulty, which is less resource demanding than the complete noise model. However, performing error mitigation for a sparse noise model may introduce more bias since second or higher order error may affect the result significantly when circuit size becomes sufficiently large. We leave a detailed analysis of the trade-off between efficiency and accuracy of using sparse noise model in future work.

During the preparation of this work, we become aware of an independent work that combines symmetry verification with probabilistic error cancellation to create the subspace noise tailoring algorithm~\cite{pec_sv}. Symmetry verification defines a computational subspace through a set of Pauli operators based on local fermion-to-qubit encoding, and it discards quantum noise that does not commute with any of these Pauli operators. It is obvious that symmetry verification acts similar with quantum error detection code, and probabilistic error cancellation in their protocol is also used to remove undetectable errors. However, the $[[n, n-2, 2]]$ quantum error detection codes always have two stabilizer generators and does not depend on the conserved quantities of problem Hamiltonian, so our protocol is easier to implement on a wider range of quantum algorithm, while their method may gives a better performance on fermionic simulation. 

While most error mitigation techniques are designed to reduce bias in expectation value, it remains an open problem whether a hybrid approach could extend the capability of error mitigation beyond fixing noisy expectation values. This is possible because quantum error correction codes can protect arbitrary logical circuits, and probabilistic error cancellation is able to fully restore the measurement statistics in principle since it completely reverses the effect of the noise channel. However, all studies of hybrid error suppression, including our work, do not explore the potential to improve measurement statistics, even though some of these protocols are appropriate for such extension. 

\section{\label{sec:acknowledgments}Acknowledgments}
This work was supported by the U.S. Department of Energy (DOE) under Contract No.~DE-AC02-05CH11231,  through the Office of Advanced Scientific Computing  Research Accelerated Research for Quantum Computing  Program. This research used resources of the Oak Ridge Leadership Computing Facility, which is a DOE Office of Science User Facility supported under Contract No.~DE-AC05-00OR22725. This research was conducted in part using IBM Quantum Systems provided through USC’s IBM Quantum Innovation Center. We thanks Meltem Tolunay, Nate Earnest-Noble and Siyuan Niu for helpful discussion on IBM real device implementation, and Ewout van den Berg for helpful discussion on noise learning and PEC based on Pauli-Lindblad noise model. 

\bibliography{ref}

\clearpage
\appendix
\renewcommand{\thefigure}{S\arabic{figure}}
\setcounter{figure}{0}
\setcounter{section}{0}

\section{\label{app:cb}Learning the Pauli Noise via Cycle Benchmarking}
A $n$-qubits Pauli noise channel is defined as $\Lambda(\rho) = \sum_{j}c_j \mathcal{P}_j(\rho)$ where the Pauli error rates $\boldsymbol{c} = \{c_j\}$ forms a normal probability distribution. The Pauli channel can be also written in the form of Pauli transfer matrix $\mathcal{T}$~\cite{introGST}, which is defined as
\begin{equation}
    \mathcal{T}_{ab} = \frac{1}{2^n}{\rm Tr}
\left[P_{a}\Lambda(P_{b})\right]. 
\end{equation}
The Pauli transfer matrix for a $n$-qubit Pauli channel has only diagonal elements. These elements are called Pauli fidelities $\boldsymbol{f} = \{f_k\}$, where the $k$-th element is defined by $f_k= {\rm Tr}\left[ P_k \Lambda(P_k) \right] / 2^n$. The transformation of Pauli fidelities $\boldsymbol{f}$ and Pauli error rates $\boldsymbol{c}$ is given by $\boldsymbol{f} = \boldsymbol{W}\boldsymbol{p}$, or 
\begin{equation}\label{eq:walsh_hadamard}
    f_k = \sum_{j}(-1)^{\langle P_j, P_k\rangle} c_j, \;c_j = \frac{1}{4^n}\sum_{k}(-1)^{\langle P_j, P_k\rangle}f_k.
\end{equation}
Here $\langle P_j, P_k\rangle$ denotes the symplectic inner product of Pauli operators, which is zero if two operators commute or one if they anti-commute. Such a transformation is referred to as Walsh-Hadamard transformation in the literature. 

The Walsh-Hadamard transform reveals an important property of a $n$-qubit Pauli noise channel $\Lambda$: each of $4^n$ Pauli operators $P_k$ is an eigen-operator of channel $\Lambda$ with eigenvalue $f_k$ (so it is also called Pauli eigenvalues). This can be seen from following equation, 
\begin{equation}\label{eq:pauli_eigenvalue}
    \Lambda(P_k) = \sum_{j}c_jP_jP_kP_j = \sum_{j}(-1)^{\langle P_j, P_k\rangle} c_j P_k = f_k P_k,
\end{equation}

\begin{figure} 
    \centering
    \includegraphics[width=\textwidth]{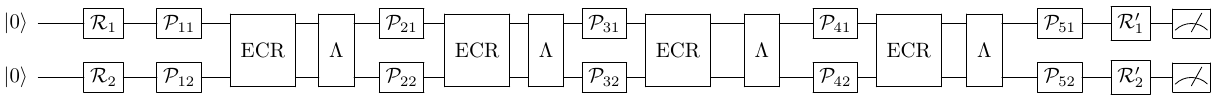}
\caption{\label{fig:cb_circuit} Cycle benchmarking circuit with four $\rm ECR$s. Here $\mathcal{R}_{1, 2}$ and $\mathcal{R}'_{1, 2}$ rotate $|00\rangle$ state to appropriate basis. $\mathcal{P}_{j1}$ and $\mathcal{P}_{j2}$ are random Pauli gates used for Pauli twirling on the first and second qubits, respectively. }
\end{figure}

Cycle benchmarking and its variants can be used to efficiently characterize Pauli channel information \cite{CB2019}. To be specific, one can directly measure the Pauli fidelities $\boldsymbol{f}$ with cycle benchmarking, then derive Pauli error rates $\boldsymbol{c}$ from Walsh-Hadamard transformation. To understand the process of learning Pauli fidelities, here we provide an example of learning individual $f_{ZI}$ of $\rm ECR$ gate noise using cycle benchmarking circuit in Fig.~\ref{fig:cb_circuit}. At the beginning, the circuit prepares the $+1$ eigenstate of operator $ZI$ with both $\mathcal{R}_1$ and $\mathcal{R}_2$ as identity. If tracing the state with its stabilizer, the first set of random Pauli gates $\mathcal{P}_{11}$ and $\mathcal{P}_{12}$ will add a phase factor of $\pm 1$ to the stabilizer. It will still be $ZI$ after propagating through the first $\rm ECR$ gate, then becomes $f_{ZI}\cdot ZI$ according to eq.~\eqref{eq:pauli_eigenvalue} after applying the noise channel (we ignore the $\pm 1$ phase here). After four noisy $\rm ECR$ gates the stabilizer becomes $(f_{ZI})^4 ZI$. At the end of the circuit, we measure operator $ZI$ and compute ${\rm Tr}[ZI\cdot\widetilde{\mathcal{U}}(\rho)] = {\rm Tr}[ZI\cdot (f_{ZI}^4 ZI \rho)] = A_{ZI}f_{ZI}^4$, where $\widetilde{\mathcal{U}}$ denotes the circuit with four noisy $\rm ECR$ gates and five sets of random Pauli. The coefficient $A_{ZI}$ is related to state preparation and measurement error, but it is independent with depth of cycle benchmarking circuit. Finally, one repeat above experiment with $m$ repetition of $\rm ECR$ for $f^{m}_{ZI}$, and perform an exponential fitting ${\rm CB}(m) = Af_{ZI}^{m}$ to get SPAM-error free Pauli fidelity $f_{ZI}$.

However, some Pauli fidelities cannot be learned from cycle benchmarking due to the existence of degeneracy~\cite{PauliNoiselearnability, pauli_lindblad}. The issue of unlearnability can be seen from below example. Suppose we are going to learn the Pauli fidelity $f_{XI}$ with top circuit in Fig.~\ref{fig:cb_circuit}. The circuit prepares the $+1$ eigenstate of $XI$ with $\mathcal{R}_1$ as Hadamard gate. Then stabilizer $XI$ will change to $YX$ after the first $\rm ECR$ gate, and become $f_{YX}\cdot YX$ after passing through the noise channel. The second ideal $\rm ECR$ will convert $YX$ back to $XI$, and the noise channel of the second $\rm ECR$ add a coefficient $f_{XI}$ to the stabilizer, which gives $f_{XI}f_{YX}XY$. Repeating this process one eventually obtain the stabilizer $(f_{XI}f_{YX})^2 XY$, which lead to $A_{XI, YX}(f_{XI}f_{YX})^2 $ as the expectation value of $XI$. Here, the Pauli fidelity $f_{XI}$ cannot be isolated from aforementioned measurement result, so we say the Pauli fidelity $f_{XI}$ is unlearnable.

\begin{figure}
    \centering
    \includegraphics[width=\textwidth]{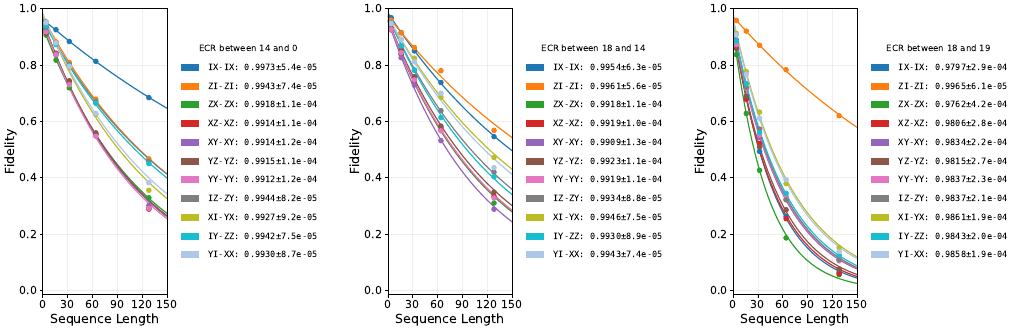}
\caption{\label{fig:fidelity_decay} The exponential fitting and results of Pauli fidelities for $\rm ECR$ gate on qubit pair with label \texttt{14-0}, \texttt{18-14} and \texttt{18-19} in \texttt{ibm\_brussels}. If the result is purely for one Pauli fidelity, e.g., $f_{IX}$, we label the results as ``IX-IX''. Here, the qubit pair shows the default ``direction'' of $\rm ECR$ gates in the given device. If there is degeneracy, e.g., $\sqrt{f_{XI}f_{YX}}$, we label the results as ``XI-YX''. This result is used for the QEDC+PEC protocol in the real device VQE experiment. }
\end{figure}

Following the discussion in Ref.~\cite{PauliNoiselearnability}, the Pauli fidelity of $IX$, $XY$, $XZ$, $YY$, $YZ$, $ZI$ and $ZX$ of Pauli noise channel of $\rm ECR$ gate are learnable via cycle benchmarking, while the degeneracy issue of $f_{IZ}f_{ZY}$, $f_{XI}f_{YX}$, $f_{IY}f_{ZZ}$ and $f_{YI}f_{XX}$ cannot be solved by any quantum experiment, even using interleaved gates $\sqrt{Z}$ and $\sqrt{X}$~\cite{PauliNoiselearnability} before each $\rm ECR$ gates. An intuition of the unlearnability is that one cannot recover identity operator in the stabilizer before or after $\rm ECR$s. In this paper, we assume the Pauli fidelity of each unlearnable pair is equal, even though they can be any possible value (e.g., $\alpha f_{IZ}$ and $f_{ZY}/\alpha$) as long as their product is equal to the measurement. 

In our real device demonstration, we run the cycle benchmarking experiment on a linear configuration of qubits with label as \texttt{19-18-14-0} on the \texttt{ibm\_brussels}. During the experiment, all $\rm ECR$s are assumed to have identical noise model if they act on same controlled-target qubit pair. Following open-source code from Ref.~\cite{PauliNoiselearnability}, we build up cycle benchmarking circuit for basis of following nine operators, $\{XX, XY, XZ, YX, YY, YZ, ZX, ZY, ZZ\}$, on each qubit pair with $\rm ECR$ sequences length as $4, 16, 32, 64, 128$. We execute individual circuit under optimization level as $1$ on the IBM device and compute ${\rm Tr}\left[P_k \,\widetilde{\mathcal{U}}(\rho)\right]$ for each of nine Pauli operators. For those basis $IX$ and $XI$, we use the result from circuits preparing the basis of $XX$ to compute the expectation value, since $IX$ and $XI$ are also stabilizer of $+1$ eigenstate of $XX$. Similarly, we measure the expectation of $IY$ from result of circuit preparing $XY$, $IZ$ from $XZ$, $YI$ from $YX$ and $ZI$ from $ZX$. Eventually, we perform an exponential fitting ${\rm CB}(m) = Af_k^{m}$ for each Pauli basis, where $m$ is the length of $\rm CNOT$ sequences, and obtain a SPAM-free Pauli fidelity $f_k$. The experimental results are shown in Fig.~\ref{fig:fidelity_decay}. Note that $f_k$ is already free from readout error so it is not necessary to apply the measurement error mitigation on benchmarking results. The Pauli error rate $\boldsymbol{c}$ is then derived from the Pauli fidelities $\boldsymbol{f}$ using the Walsh-Hadamard transformation, which is not shown here but used for the noise estimation in VQE experiment discussed in Sec.~\ref{sec:ibm}.

\section{\label{app:rot} Error Propagation with Rotation}
Consider the error propagation identity $\mathcal{V}\circ\mathcal{P} = \mathcal{M}\circ\mathcal{V}$ shown in eq.~\eqref{eq:err_prop}, where $\mathcal{V}(\rho) = V\rho V^{\dagger}$ and $\mathcal{M}(\rho) = M\rho M^{\dagger}$. If $V$ is a rotation gate, then $M$ can be derived by following relations. For $R_x(\theta)$ gate,
\begin{align}\label{eq:rx_prop}
    R_x(\theta)XR_x^{\dagger}(\theta) =& X, \\
    R_x(\theta)YR_x^{\dagger}(\theta) =& \cos\theta Y + \sin\theta Z, \\
    R_x(\theta)ZR_x^{\dagger}(\theta) =& \cos\theta Z - \sin\theta Y.
\end{align}
For $R_y(\theta)$ gate,
\begin{align}\label{eq:ry_prop}
    R_y(\theta)XR_y^{\dagger}(\theta) =& \cos\theta X - \sin\theta Z, \\
    R_y(\theta)YR_y^{\dagger}(\theta) =& Y, \\
    R_y(\theta)ZR_y^{\dagger}(\theta) =& \cos\theta Z+ \sin\theta X.
\end{align}
For $R_z(\theta)$ gate, 
\begin{equation}\label{eq:rz_prop}
    \begin{split}
        R_z(\theta)XR_z^{\dagger}(\theta) =& \cos\theta X + \sin\theta Y, \\
        R_z(\theta)YR_z^{\dagger}(\theta) =& \cos\theta Y - \sin\theta X, \\
      R_z(\theta)ZR_z^{\dagger}(\theta) =& Z. 
    \end{split}
\end{equation}

If $M$ is a linear combination of Pauli operator, the channel $\mathcal{M}$ has non-zero off-diagonal terms in its $\chi$-matrix representation, $\mathcal{M}(\rho) = \sum_{j,k}\chi_{jk}P_{j}\rho P_k$, so $\mathcal{M}$ is no longer a Pauli channel. This can be seen from an example of $X$ error after propagating through $R_z(\theta)$ gate, where the error becomes $M = \cos\theta X + \sin\theta Y$ and the channel below is not a Pauli channel,
\begin{equation}\label{eq:example}
    \begin{split}
    \mathcal{M} =& (\cos\theta X + \sin\theta Y)\rho (\cos\theta X + \sin\theta Y)^{\dagger} \\
    =& \cos^2\theta X\rho X + \cos\theta\sin\theta (X\rho Y + Y\rho X) + \sin^2\theta Y\rho Y.
    \end{split}
\end{equation}

According to the discussion of eq.~\eqref{eq:coherent_error}, the $[[n, n-2, 2]]$ code is able to reduce noise in the form of $\mathcal{N}(\rho) = \sum_s q_s E_s\rho E_s$, where $E_s = \sum_{s} a_{r_s}P_s$ could be a linear combination of Pauli and $\{a^2_{r_s}\}$ gives a probability distribution. For example, the noise in Eq.~\eqref{eq:example} can be fully removed by the QEDC since both $X$ and $Y$ anti-commute with its $Z$-type stabilizer generator. However, in general, QEDC cannot eliminate all possible errors. In such cases, the noisy expectation value is given by
\begin{equation}\label{eq:noisy_exp_appendix}
    {\rm Tr}\left[A\,\mathcal{N}_{\rm logical}(\rho) \right] = \sum_{s, r_s} q_s a^2_{r_s} {\rm Tr}\left[ A\, P_{r_s}\rho P_{r_s} \right] + \sum_{\substack{s, r_s, k_s\\ r_s\neq k_s}} q_s a_{r_s}a_{k_s} {\rm Tr}\left[ A\, P_{r_s}\rho P_{k_s} \right].
\end{equation}
It is resource consuming to completely recover above expectation value via PEC, which will require $16^n$ basis operations to implement the inverse of a $n$-qubit logical error channel~\cite{endo2018practical}. Instead, one can approximate the noiseless expectation value by neglecting the impact from off-diagonal terms and cancelling only the diagonal part of the error channel, i.e., $\mathcal{N}_{\rm reduced}(\rho) = \sum_s q_s a^2_{r_s} P_{r_s}\rho P_{r_s}$, which only requires at most $4^n$ basis operations to implement the Pauli channel $\mathcal{N}^{-1}_{\rm reduced}(\rho)$.

It is worth mentioning that the contribution of some off-diagonal terms to the noisy expectation value can be neglected. We can rewrite each term in the second summation of eq.~\eqref{eq:noisy_exp_appendix} as follows,
\begin{equation}
    \begin{split}
        {\rm Tr}\left[ A\, P_{r_s}\rho P_{k_s} \right] =& {\rm Tr}\left[ P_{k_s} A\, P_{r_s}\rho  \right] \\
        =& {\rm Tr}\left[(-1)^{\langle P_{k_s}, A \rangle}(-1)^{\langle P_{k_s}, P_{r_s} \rangle}(-1)^{\langle P_{r_s}, A \rangle} P_{r_s} A\,P_{k_s} \rho \right] \\
        =& {\rm Tr}\left[(-1)^{\langle P_{k_s}, A \rangle}(-1)^{\langle P_{k_s}, P_{r_s} \rangle}(-1)^{\langle P_{r_s}, A \rangle} A\,P_{k_s} \rho  P_{r_s} \right]
    \end{split}
\end{equation}
where $\langle P_A, P_B\rangle$ is sympletic inner product of two Pauli operators $P_A$ and $P_B$. In this context, the off-diagonal terms $ {\rm Tr}\left[ A\, P_{r_s}\rho P_{k_s} \right]$ and $ {\rm Tr}\left[ A\, P_{k_s}\rho P_{r_s}\right]$ in eq.~\eqref{eq:noisy_exp_appendix} will cancel if $P_{k_s}A\, P_{r_s}$ anticommutes with $P_{r_s}A\, P_{k_s}$. Moreover, the noisy expectation may not be significantly affected by remaining off-diagonal part when the error rate $q_s$ is very small. Therefore, we believe that inversing $\mathcal{N}_{\rm reduced}$ will possibly gives a good approximation of ideal expectation value. 

For the purpose of illustration, we explain how the idea of canceling $\mathcal{N}_{\rm reduced}$ works on a single-qubit example. Suppose $|\psi\rangle = \alpha |0\rangle + \beta|1\rangle$ is a noiseless single-qubit state, then the ideal expectation value of observable $Z$ is given by $\langle \psi| Z|\psi\rangle = |\alpha|^2 - |\beta|^2$. When an error $M = \cos\theta X + \sin\theta Y$ occurs, according to Eq.~\eqref{eq:example}, the noisy expectation value is derived by
\begin{equation}\label{eq:true_trace}
    \begin{split}
        {\rm Tr}\left[Z\,\mathcal{M}(\rho)\right] =& \cos^2\theta {\rm Tr}\left[ZX\rho X\right]  + \cos\theta\sin\theta {\rm Tr}\left[ZX\rho Y\right]\\
        &\qquad\qquad\qquad+\cos\theta\sin\theta {\rm Tr}\left[ZY\rho X\right] + \sin^2\theta {\rm Tr}\left[ZY\rho Y\right] \\
        =& -|\alpha|^2 + |\beta|^2 +\cos\theta\sin\theta {\rm Tr}\left[ZX\rho Y\right] +\cos\theta\sin\theta {\rm Tr}\left[ZY\rho X\right] \\
        =& -|\alpha|^2  + |\beta|^2.
    \end{split}
\end{equation}
Here, $XZY = -YZX$ so the off-diagonal terms in the second equality vanish, and the noisy expectation value is equivalent with the one under $\mathcal{N}_{\rm reduced}(\rho) = \cos^2\theta X\rho X + \sin^2\theta Y\rho Y$, where
\begin{equation}
    {\rm Tr}\left[Z\,\mathcal{N}_{\rm reduced}(\rho)\right] = \cos^2\theta {\rm Tr}\left[ZX\rho X\right] + \sin^2\theta {\rm Tr}\left[ZY\rho Y\right] = -|\alpha|^2 + |\beta|^2.
\end{equation}
In this example, the ideal expectation value can be recovered by implementing $\mathcal{N}^{-1}_{\rm reduced}(\rho)$ via quasi-probability sampling. 

\section{\label{app:sampling} Sampling Overhead of our Protocol}
In our toy model for evaluation and comparison, we consider $n$-qubits depolarizing noise~\cite{Nielsen_Chuang_2010} for each layer, 
\begin{equation}
    \mathcal{N}(\rho) = (1-p)\rho + p\frac{I}{2^n},
\end{equation}
where $I/2^n$ is the completely mixed state. The Pauli transfer matrix of above channel is a $4^n\times 4^n$ diagonal matrix given by
\begin{equation}
    R_{\mathcal{N}} = {\rm diag}(1, 1-p, \dotsc, 1-p).
\end{equation}

\begin{figure}
    \raggedright
    \includegraphics[width=0.95\textwidth]{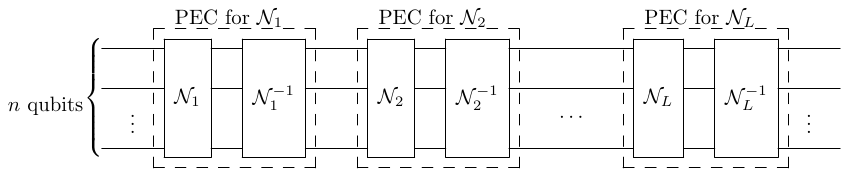}\\
    \vspace{10pt}
    \includegraphics[width=0.65\textwidth]{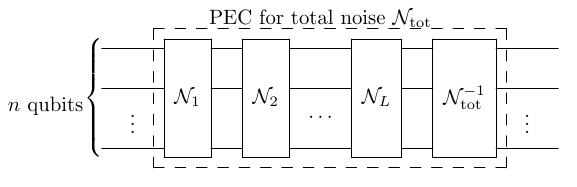}\\
    \vspace{10pt}
    \includegraphics[width=0.8\textwidth]{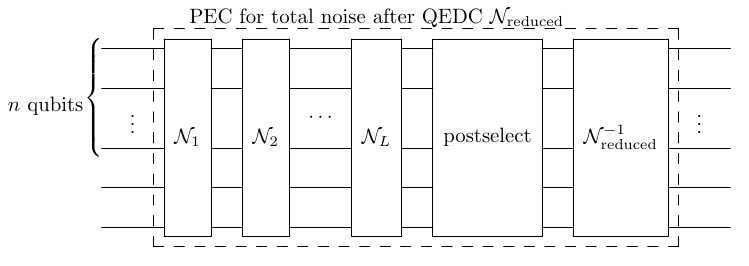}
\caption{\label{fig:sampling} Three protocols for overhead analysis. The top one is for the regular PEC. The middle one is for PEC canceling overall noise. The bottom one is for our combined protocol with QEDC and PEC. }
\end{figure}

Here we consider three settings of error mitigation in the numerical test. The first protocol is regular PEC protocol that cancel noise after each layer. The second protocol is PEC that cancel estimated overall noise. Although the idea of cancelling overall noise is proposed in Sec.~\ref{sec:noise_estimation} for our combined protocol, it can be still applied independently. The third protocol is the one proposed in Sec.~\ref{sec:protocol}. 

We then compute the sampling overhead for each protocol on a $n=4$ qubits circuit without any unitary operation (see Fig.~\ref{fig:sampling}). For the noise estimation in the second and third protocol, we use a property that the Pauli transfer matrix of channel $\mathcal{N}_2\circ \mathcal{N}_{1}$ is the product of Pauli transfer matrices of the two channels, i.e., $R_{\mathcal{N}_2\circ \mathcal{N}_{1}} = R_{\mathcal{N}_2}R_{\mathcal{N}_{1}}$~\cite{introGST}. For the last protocol, we check and remove all Pauli terms in total noise that anticommute with stabilizer generators $X^{\otimes n}$ and $Z^{\otimes n}$ for post-selection, and renormalized the coefficient to obtain $\mathcal{N}_{\rm reduced}$. Note that all noise channels, including $\mathcal{N}_{\rm tot}$ and $\mathcal{N}_{\rm reduced}$, are Pauli channels, and their inverse, $\mathcal{N}^{-1} = \sum_j c_j^{\rm inv} P_j\rho P_j$, can be computed by Walsh-Hadamard transform~\cite{pauli_lindblad}
\begin{equation}
    c_j^{\rm inv} = \frac{1}{4^n}\sum_k (-1)^{\langle P_j, P_k\rangle} \frac{1}{f_k},
\end{equation}
where $f_k$ is the Pauli fidelity of noise channel $\mathcal{N}$. Finally, one would have $\gamma = \sum_j |c_j^{\rm inv}|$ for each inverse channel, and $\gamma^2$ for sampling overhead. 

\section{\label{app:partial} Additional Cycle Benchmarks}

In this section we provide additional cycle benchmarking results for native multi-qubit unitaries on the IBM quantum platform. In case case we apply no twirling, Pauli twirling, and partial Pauli twirling, in combination with a $3$-qubit bit-flip code as defined in Section \ref{sec:QEC}.

For the $CZ$ gate, we benchmark the noisy unitary $\Tilde{\mathcal{U}} = CZ_{1,2}CZ_{2,3}$.
\begin{figure}
    \centering
    \includegraphics[width=0.45\linewidth]{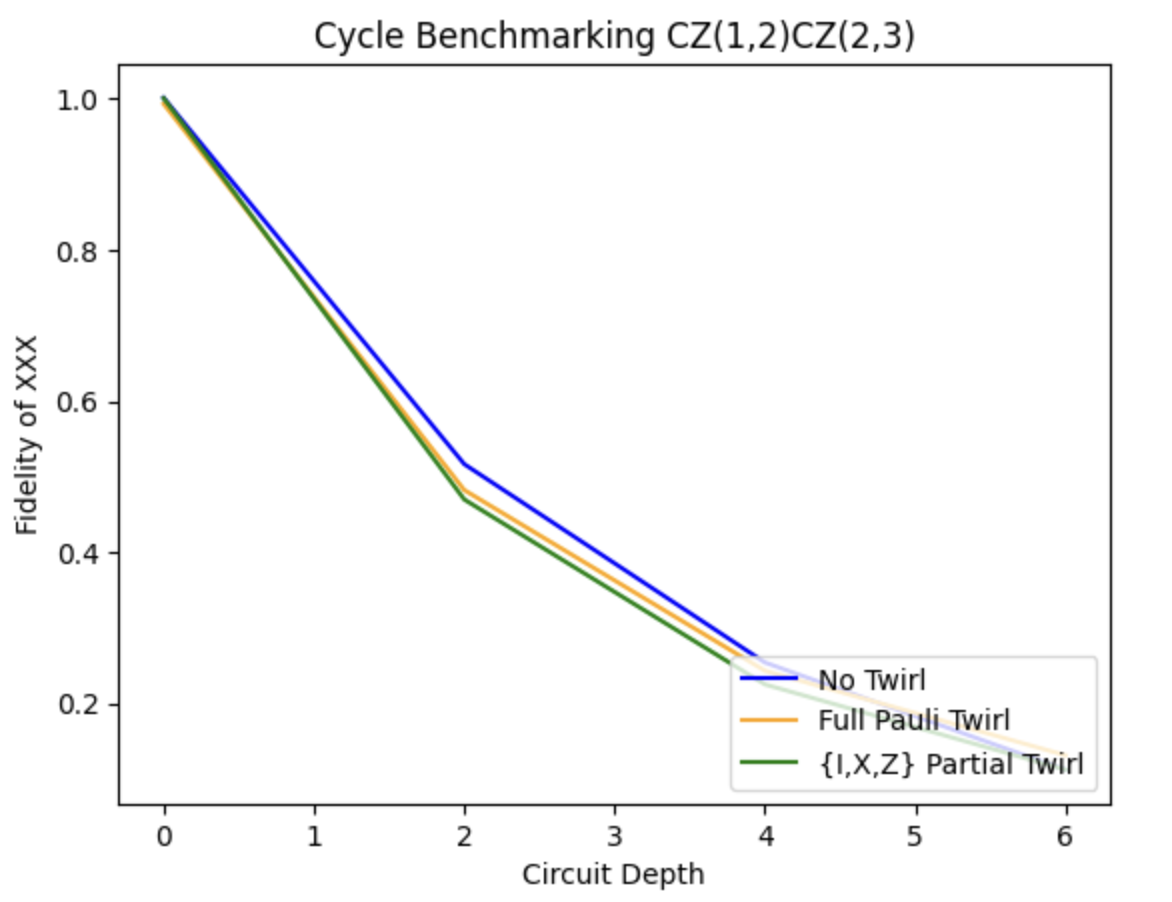}
    \includegraphics[width=0.45\linewidth]{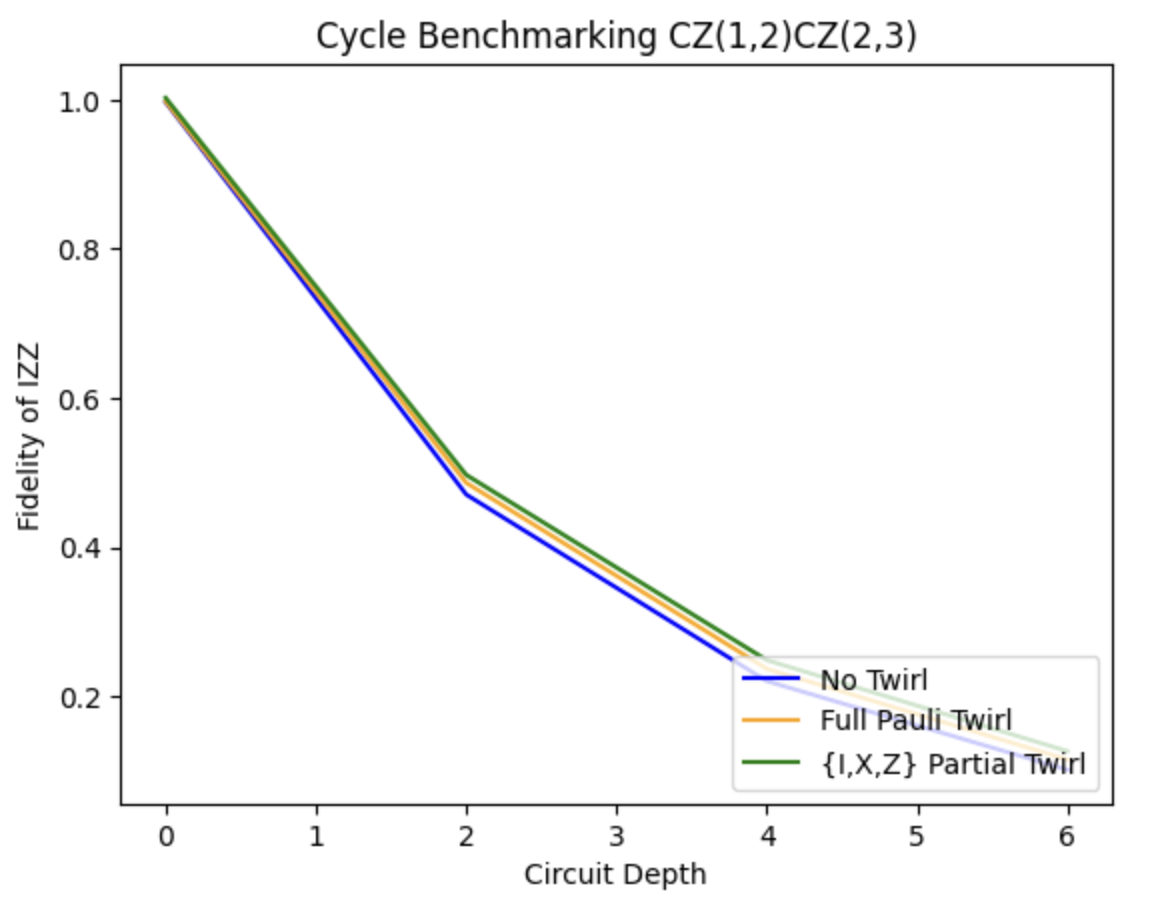}
\caption{Cycle benchmarking performed on $\Tilde{\mathcal{U}} = CZ_{1,2}CZ_{2,3}$ using twirling with bit-flip error correction. Experiment conducted on ibm brussels device. The fidelity of $GHZ_3$ stabilizers $XXX$ and $IZZ$ are plotted against increasing twirled layers of $\Tilde{\mathcal{U}}$. For partial twirling, the twirling gates are selected to maximize $X$ error propagation, and minimize all others. Performances compared between no twirling, full Pauli twirling using $I,X,Y,Z$ gates, and partial twirling using $I,X,Z$ gates.}
\label{fig:cx_cyclebench}
\end{figure}

For the $ECR$ gate, we benchmark the noisy unitary $\Tilde{\mathcal{U}} = ECR{1,2}ECR{2,3}$.
\begin{figure}
    \centering
    \includegraphics[width=0.45\linewidth]{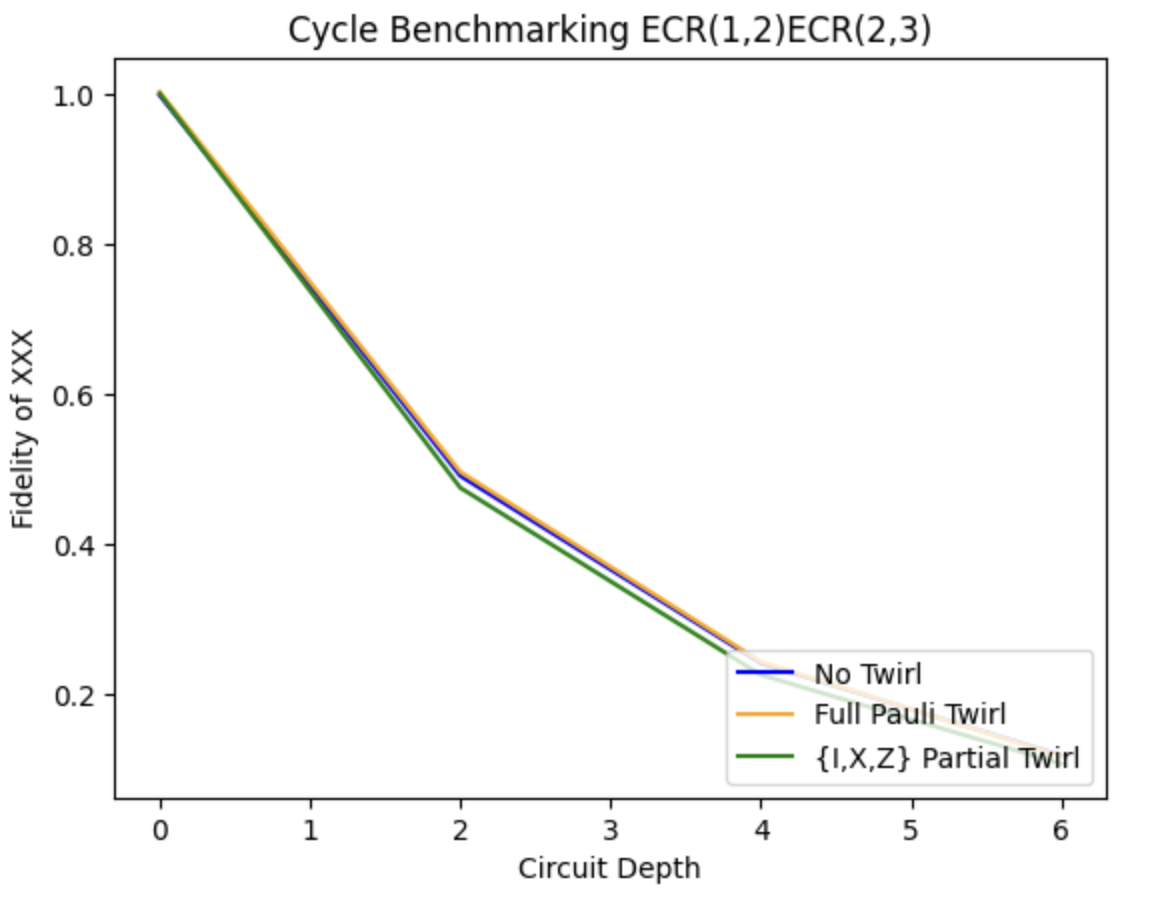}
    \includegraphics[width=0.45\linewidth]{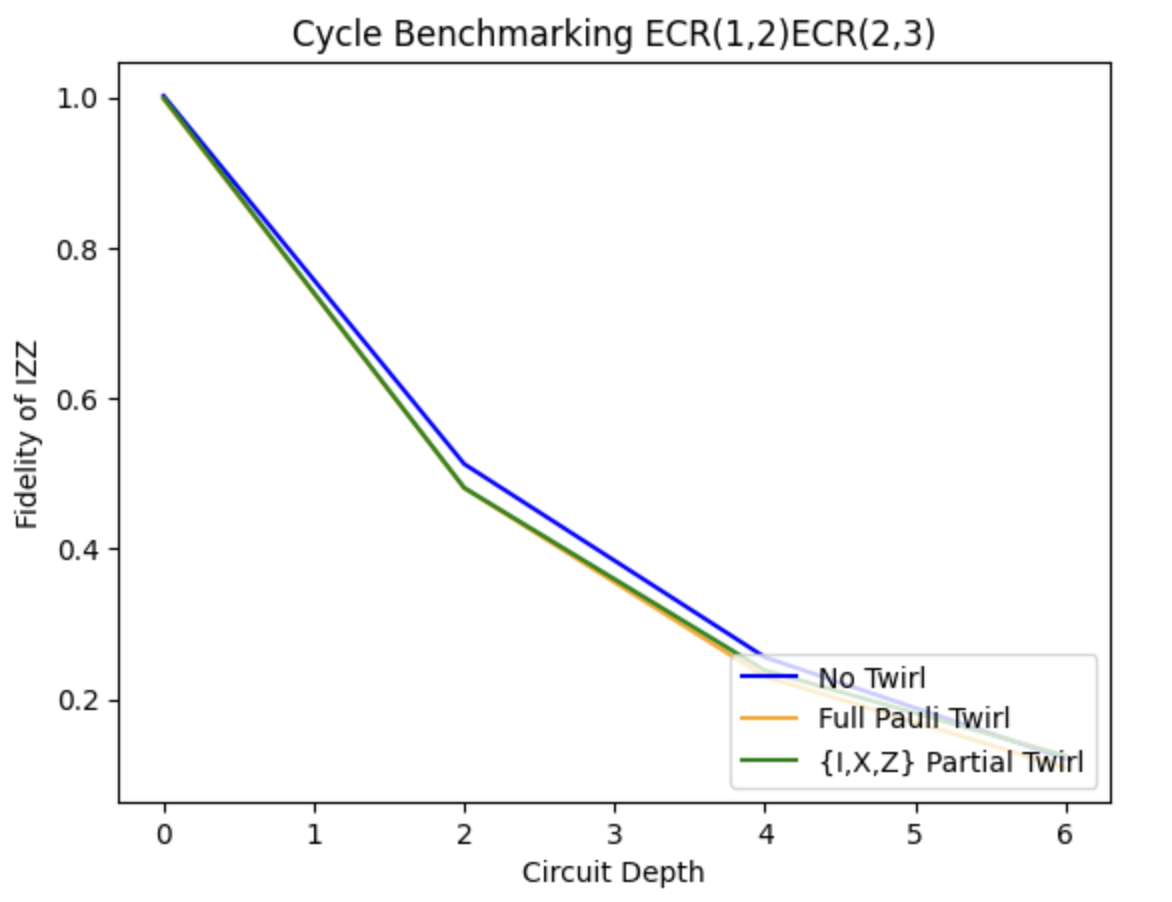}
\caption{Cycle benchmarking performed on $\Tilde{\mathcal{U}} = ECR_{1,2}ECR_{2,3}$ using twirling with bit-flip error correction. Experiment conducted on ibm brussels device. The fidelity of $GHZ_3$ stabilizers $XXX$ and $IZZ$ are plotted against increasing twirled layers of $\Tilde{\mathcal{U}}$. For partial twirling, the twirling gates are selected to maximize $X$ error propagation, and minimize all others. Performances compared between no twirling, full Pauli twirling using $I,X,Y,Z$ gates, and partial twirling using $I,X,Z$ gates.}
\label{fig:cx_cyclebench}
\end{figure}

For the $iSWAP$ gate, we benchmark the noisy unitary $\Tilde{\mathcal{U}} = iSWAP_{1,2}iSWAP_{2,3}$.
\begin{figure}
    \centering
    \includegraphics[width=0.45\linewidth]{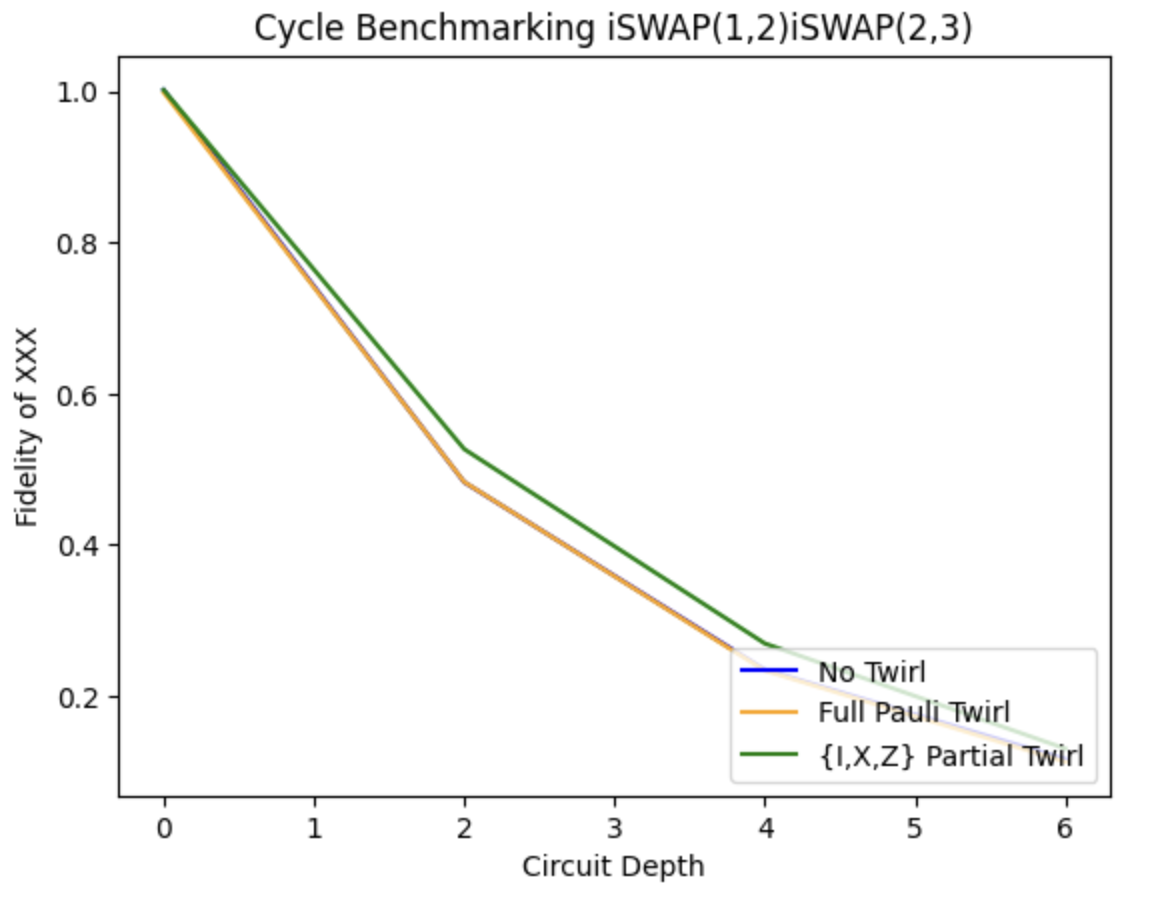}
    \includegraphics[width=0.45\linewidth]{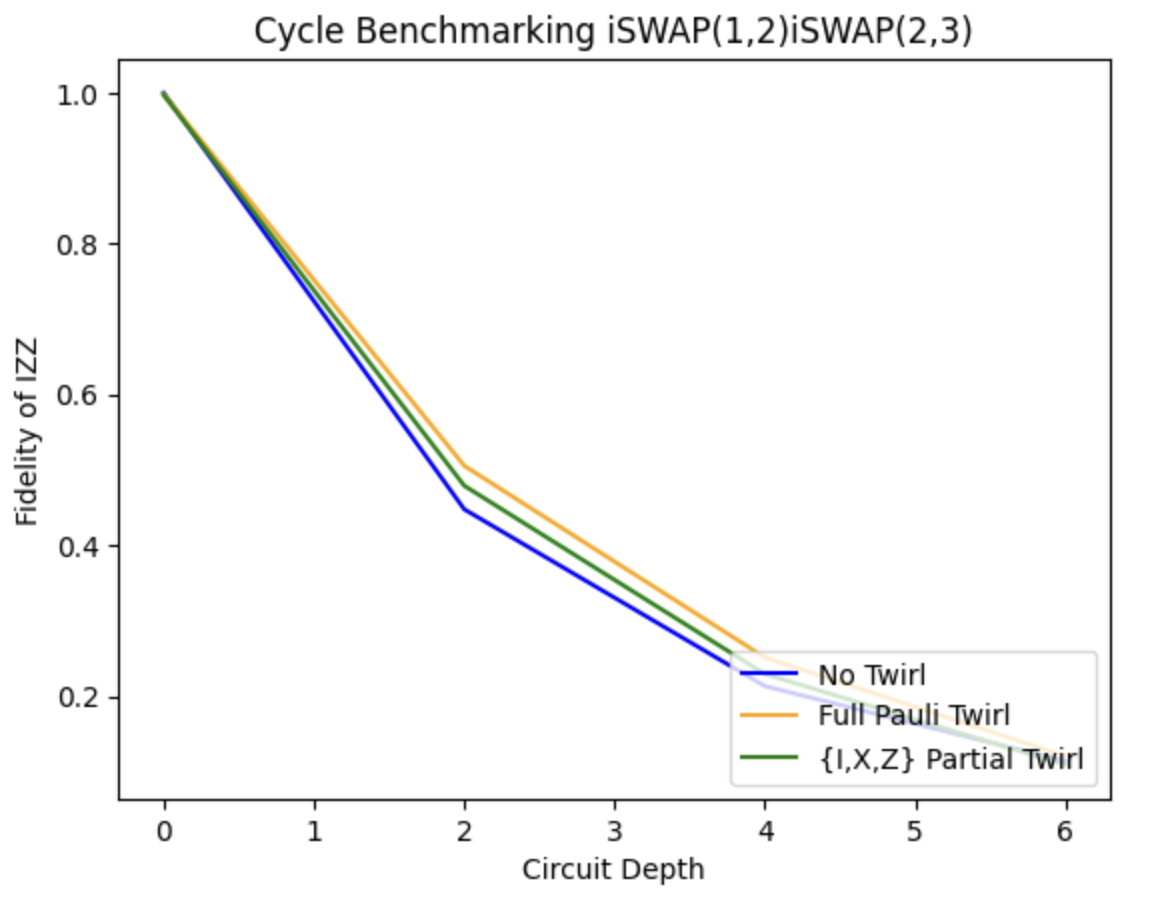}
\caption{Cycle benchmarking performed on $\Tilde{\mathcal{U}} = iSWAP_{1,2}iSWAP_{2,3}$ using twirling with bit-flip error correction. Experiment conducted on ibm brussels device. The fidelity of $GHZ_3$ stabilizers $XXX$ and $IZZ$ are plotted against increasing twirled layers of $\Tilde{\mathcal{U}}$. For partial twirling, the twirling gates are selected to maximize $X$ error propagation, and minimize all others. Performances compared between no twirling, full Pauli twirling using $I,X,Y,Z$ gates, and partial twirling using $I,X,Z$ gates.}
\label{fig:cx_cyclebench}
\end{figure}

\section{\label{app:coeffient}Coefficients for Hamiltonian}
Table~\ref{tab:coefficient} shows the coefficients for Eq.~\eqref{eq:energy} as a function of the internuclear separation $R$ from Ref.~\cite{coefficient_H2}. The electronic integrals are calculated using in the STO-3G (GTO) basis set.

\begin{table*}[htp]
    \centering
    \caption{}\label{tab:coefficient}
    \begin{tabular}{c@{\hspace{15pt}}c@{\hspace{15pt}}c@{\hspace{15pt}}c@{\hspace{15pt}}c@{\hspace{15pt}}c}
        \hline\hline
        $R$ (\AA) & $g_1$ & $g_2$ & $g_5$ & $g_3$ & $g_4$  \\
        \hline
        0.05 & 10.0777 & -1.05533 & 0.155708 & -1.05533 & 0.0139333 \\
        0.1 & 4.75665 & -1.02731 & 0.15617 & -1.02731 & 0.0138667 \\
        0.15 & 2.94817 & -0.984234 & 0.15693 & -0.984234 & 0.013761 \\
        0.2 & 2.01153 & -0.930489 & 0.157973 & -0.930489 & 0.0136238 \\
        0.25 & 1.42283 & -0.870646 & 0.159277 & -0.870646 & 0.0134635 \\
        0.3 & 1.01018 & -0.808649 & 0.160818 & -0.808649 & 0.013288 \\
        0.35 & 0.701273 & -0.747416 & 0.162573 & -0.747416 & 0.0131036 \\
        0.4 & 0.460364 & -0.688819 & 0.164515 & -0.688819 & 0.012914 \\
        0.45 & 0.267547 & -0.63389 & 0.166621 & -0.63389 & 0.0127192 \\
        0.5 & 0.110647 & -0.58308 & 0.16887 & -0.58308 & 0.0125165 \\
        0.55 & -0.0183734 & -0.536489 & 0.171244 & -0.536489 & 0.0123003 \\
        0.65 & -0.213932 & -0.455433 & 0.176318 & -0.455433 & 0.0118019 \\
        0.75 & -0.349833 & -0.388748 & 0.181771 & -0.388748 & 0.0111772 \\
        0.85 & -0.445424 & -0.333747 & 0.187562 & -0.333747 & 0.0104061 \\
        0.95 & -0.513548 & -0.287796 & 0.19365 & -0.287796 & 0.00950345 \\
        1.05 & -0.5626 & -0.248783 & 0.199984 & -0.248783 & 0.00850998 \\
        1.15 & -0.597973 & -0.215234 & 0.206495 & -0.215234 & 0.00747722 \\
        1.25 & -0.623223 & -0.186173 & 0.213102 & -0.186173 & 0.00645563 \\
        1.35 & -0.640837 & -0.160926 & 0.219727 & -0.160926 & 0.00548623 \\
        1.45 & -0.652661 & -0.138977 & 0.226294 & -0.138977 & 0.0045976 \\
        1.55 & -0.660117 & -0.119894 & 0.23274 & -0.119894 & 0.00380558 \\
        1.65 & -0.664309 & -0.103305 & 0.239014 & -0.103305 & 0.00311545 \\
        1.75 & -0.666092 & -0.0888906 & 0.245075 & -0.0888906 & 0.0025248 \\
        1.85 & -0.666126 & -0.0763712 & 0.250896 & -0.0763712 & 0.00202647 \\
        1.95 & -0.664916 & -0.0655065 & 0.256458 & -0.0655065 & 0.001611 \\
        2.05 & -0.662844 & -0.0560866 & 0.26175 & -0.0560866 & 0.00126812 \\
        2.15 & -0.660199 & -0.0479275 & 0.266768 & -0.0479275 & 0.000988 \\
        2.25 & -0.657196 & -0.0408672 & 0.271512 & -0.0408672 & 0.000761425 \\
        2.35 & -0.653992 & -0.0347636 & 0.275986 & -0.0347636 & 0.000580225 \\
        2.45 & -0.650702 & -0.0294924 & 0.280199 & -0.0294924 & 0.000436875 \\
        2.55 & -0.647408 & -0.0249459 & 0.28416 & -0.0249459 & 0.000325025 \\
        2.65 & -0.644165 & -0.0210309 & 0.287881 & -0.0210309 & 0.0002388 \\
        2.75 & -0.641011 & -0.0176672 & 0.291376 & -0.0176672 & 0.0001733 \\
        2.85 & -0.637971 & -0.0147853 & 0.294658 & -0.0147853 & 0.0001242 \\
        2.95 & -0.635058 & -0.0123246 & 0.297741 & -0.0123246 & 8.7875e-05 \\
        3.05 & -0.632279 & -0.0102318 & 0.300638 & -0.0102317 & 6.145e-05 \\
        3.15 & -0.629635 & -0.00845958 & 0.303362 & -0.00845958 & 4.2425e-05 \\
        3.25 & -0.627126 & -0.00696585 & 0.305927 & -0.00696585 & 2.895e-05 \\
        3.35 & -0.624746 & -0.0057128 & 0.308344 & -0.0057128 & 1.955e-05 \\
        3.45 & -0.622491 & -0.0046667 & 0.310625 & -0.0046667 & 1.305e-05 \\
        3.55 & -0.620353 & -0.00379743 & 0.31278 & -0.00379743 & 8.575e-06 \\
        3.65 & -0.618325 & -0.0030784 & 0.314819 & -0.0030784 & 5.6e-06 \\
        3.75 & -0.616401 & -0.00248625 & 0.31675 & -0.00248625 & 3.6e-06 \\
        3.85 & -0.614575 & -0.00200063 & 0.318581 & -0.00200062 & 2.275e-06 \\
        3.95 & -0.612846 & -0.00160393 & 0.32032 & -0.00160392 & 1.425e-06 \\
        \hline
    \end{tabular}
\end{table*}

\end{document}